\documentclass[english,aps, prx, reprint, superscriptaddress,longbibliography]{revtex4-1}
\usepackage[T1]{fontenc}
\usepackage[utf8]{inputenc}
\usepackage{amsmath}
\usepackage{amssymb}
\usepackage{graphicx}
\usepackage{physics}

\makeatletter
\usepackage{hyperref}
\usepackage{babel}
\hypersetup{
	colorlinks = true,
	allcolors = blue
}

\makeatother

\usepackage{babel}

\newcommand{\PKU}{International Center for Quantum Materials and School of Physics, Peking University, Beijing 100871, China}
\newcommand{\SIQSE}{Shenzhen Institute for Quantum Science and Engineering, Southern University of Science and Technology, Shenzhen 518055, China}
\newcommand{\IQA}{International Quantum Academy, Shenzhen 518048, China}
\newcommand{\HFNLHF}{Hefei National Laboratory, Hefei 230088, China}
\newcommand{\GDKL}{Guangdong Provincial Key Laboratory of Quantum Science and Engineering, Southern University of Science and Technology, Shenzhen 518055, China}
\newcommand{\NKU}{ Chern Institute of Mathematics and LPMC, Nankai University, Tianjin 300071, China}
\newcommand{\NUST}{School of Mathematics and Statistics, Nanjing University of Science and Technology, Nanjing 210094, China}

\begin{document}
\title{The fundamental localization phases in quasiperiodic systems: \\
A unified framework and exact results}

\author{Xin-Chi Zhou}
\affiliation{\PKU}\affiliation{\HFNLHF}

\author{Bing-Chen Yao}
\affiliation{\PKU}

\author{Yongjian Wang}
\affiliation{\NUST}

\author{\\ Yucheng Wang} 
\affiliation{\SIQSE}\affiliation{\IQA}\affiliation{\GDKL}

\author{Yudong Wei}
\affiliation{\PKU}\affiliation{\HFNLHF}

\author{Qi Zhou}
\affiliation{\NKU}

\author{Xiong-Jun Liu}
\email{xiongjunliu@pku.edu.cn}
\affiliation{\PKU}\affiliation{\HFNLHF}\affiliation{\IQA}

\begin{abstract}
The disordered quantum systems host three classes of quantum states, the extended, localized,
and critical, which bring up seven distinct fundamental phases in nature: three pure phases and four
coexisting ones with mobility edges, yet a unified theory built on universal mechanism and full realization of all these phases has not been developed. Here we propose a unified framework based on a spinful quasiperiodic (QP) system which realizes all the fundamental localization phases, with the exact and universal results being obtained for their characterization. First, we show that the pure phases are obtained when the chiral(-like) symmetry preserves in the proposed spinful
QP model, giving a criterion for emergence of the pure phases and otherwise the coexisting
ones. Further, we uncover a novel mechanism for the critical states that their emergence is protected by the generalized incommensurate matrix element zeros in the spinful QP model, which considerably broadens rigorous realizations of the exotic critical states. We then show criteria of exact solvability for the present spinful QP system, with which we construct various exactly solvable models for all distinct localization phases. In particular, we propose two novel models, dubbed spin-selective QP lattice model and QP optical Raman lattice model, to achieve all basic types of mobility edges and all the seven fundamental phases of Anderson localization physics, respectively.
The experimental scheme is proposed and studied in detail to realize these models with high feasibility. This study establishes a complete and profound theoretical framework which enables an in-depth exploration of the broad classes of all fundamental localization phenomena in QP systems, and offers key insights for constructing their exactly solvable models with experimental feasibility.
\end{abstract}
\maketitle

\section{Introduction}

Anderson localization, wherein quantum states localize exponentially
in real space due to disorder-induced scattering, is a fundamental
and universal phenomenon in disordered systems \citep{anderson1958,lee1985,kramer1993,evers2008}.
It manifests distinctly across two categories of aperiodic systems:
disordered systems and quasiperiodic lattices. In disordered systems,
scaling theory predicts that noninteracting states are localized in
one and two dimensions even under weak disorder, whereas in three
dimensions, an Anderson transition from extended to localized states
occurs only with sufficiently strong disorder \citep{abrahams1979}.
In contrast, quasiperiodic (QP) lattices exhibit richer physics including Anderson transitions
in all spatial dimensions, governed by quasiperiodic parameters \citep{aubry1980,roati2008}.
Importantly, both extended and localized states can coexist within
a single quantum phase, separated by characteristic energies known
as mobility edges (MEs) \citep{soukoulis1982b,dassarma1988,boers2007,biddle2010,ganeshan2015,li2017,gopalakrishnanSelfdualQuasiperiodicSystems2017,an2018,luschen2018a,kohlert2019a,deng2019,wang2020a,liMobilityEdgeIntermediate2020,duthie2021,an2021,roy2021a,roy2021,wang2021c,wang2022b,longhi2023a,wangExactMobilityEdges2023,gaoCoexistenceExtendedLocalized2023,longhi2024,longhi2024a,wei2024,gao2024,liu2024j,tabanelli2024,shimasakiReversiblePhasonicControl2024}. { The ability to derive exact analytical expressions for characteristic length scales and MEs in various QP models~\cite{wang2020a,wang2021c,zhou2023c,goncalves2023b,liu2022} plays a crucial role in advancing the fundamental understanding of localization transitions, providing theoretical insights that extend beyond numerical simulation.}

Between the extended and localized states lie the critical states, which
are delocalized in both position and momentum spaces. Critical
states have recently attracted extensive research interests \citep{hatsugai1990,han1994,tanese2014b,zhou2023c,wang2016b,liu2015,zeng2016,yao2019a,wang2020b,goblot2020a,xiao2021,liu2022,li2023b,wang2022a,goncalves2023b,lin2023,leeCriticaltoinsulatorTransitionsFractality2023,dai2023,shimasaki2024b,li2024d,wang2025,huang2025,chen2024e,guo2024,dottiMeasuringLocalizationPhase2024,yang2024f,duncan2024a,yao2024a,baiTunablyPolarizedDriving2025} due to their unique properties, such as local scale invariance,
multifractality, and critical quantum dynamics. Recent advances, particularly
through Avila's global theory \citep{avila2015,zhou2023c},
{refine} a rigorous characterization~\cite{simon1989,jitomirskaya2012} of the critical states, showing
that critical states generally arise when
the hopping couplings in a QP system are non-uniform and possess incommensurately
distributed zeros (IDZs) in the thermodynamic limit~\cite{liu2024g}. The exact theory enables unambiguous determination of critical phases and the new MEs between critical and other states~\cite{zhou2023c}. {Furthermore, this rigorous characterization of critical states provides a well-defined foundation for incorporating many-body interactions, thereby extending the paradigm of criticality beyond the single-particle picture.} Inclusion
of interactions further enriches the physics of the critical phase,
with multifractality of the wave function influencing both ground
state properties associated with symmetry breaking~\citep{liu2024g,feigelman2007,burmistrov2012,zhao2019,sacepe2020,goncalves2024}
and the emergence of many-body critical (MBC) phases \citep{wang2021a,wang2020b}
at infinite temperature. The MBC phase, interpolating the thermal phase
and many-body localization (MBL) \citep{pal2010a,nandkishore2015b,schreiber2015,bordia2016,bordia2017,luschen2017,bordia2017a},
signifies a non-ergodic but delocalized regime that violates the eigenstate
thermalization hypothesis (ETH) \citep{dalessio2016,deutsch1991,rigol2008,srednicki1994}.

The recently developed quasiperiodic mosaic models \citep{wang2020a,zhou2023c,gao2024,huang2025} and their generalizations~\cite{longhi2023a,dai2023,lin2024,guo2024,wei2024,longhi2024} advance the rigorous study of localization physics in QP systems, but many important issues remain unexplored. 
The extended, localized, and critical states bring up seven fundamental localization phases in nature, including three pure phases and four coexisting ones (three doubly and one triply coexisting phases). However, thus far there was no unified quantum system which is shown to host all these phases. Especially, the exactly solvable models for the complete set of such phases are crucial for the analytical study of the generic localization physics, but are currently lacking.
{Moreover, the renormalization group theory developed in Ref.~\cite{goncalves2023a} provides important insights into the characterization of transitions between the distinct types of localization phases and mobility edges, while a unified and rigorous framework that elucidates the underlying universal physical mechanisms for all these fundamental phases remains elusive.}
Besides, while Avila's global theory enables an analytic study of the spinless models, generalization of the study to the cases beyond spinless systems, {or even pointing out the condition for the spinful case can be reduced to the effective spinless model,} is not only important in fundamental physics, but also desired in experiment. For instance, the generic mechanism for critical states in the complex systems with internal degrees of freedom,
such as spins, remains unknown. 
On the other hand, the remarkable experimental progresses in engineering the spin and orbital degrees of freedom~\cite{zhang2018,liu2014,wu2016,wang2018,sun2018,song2019,lu2020,wang2021,zhang2024e} have enabled the realization of complex QP systems beyond spinless toy models. These crucial fundamental issues motivate us to establish a generic theory
that can generically characterize {a broad class of} spinful QP lattice systems without
relying on specific microscopic details.

\begin{figure*}[t]
\includegraphics{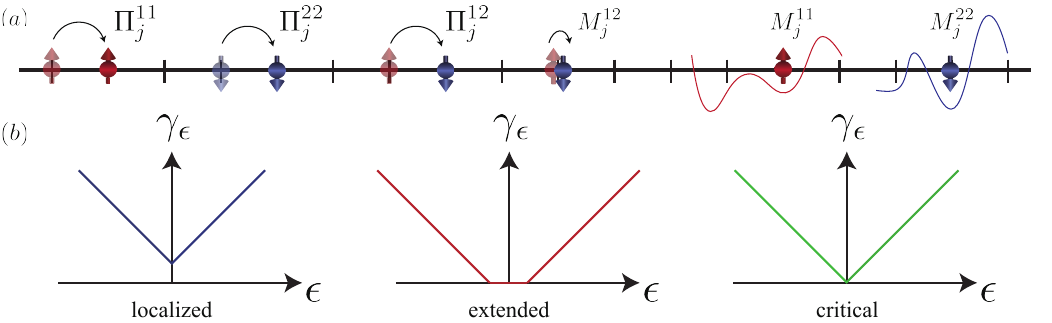}\caption{\label{fig:Fig1}Illustration of generic framework. (a) The different
processes of the spinful quasiperiodic systems. The hopping coupling
matrix $\Pi_{j}^{s,s'}$ denote the momentum transfer coupling with
the diagonal (off-diagonal) terms for the same (different) internal
degrees of freedom. The on-site matrix $M_{j}^{s,s'}$ represents
the on-site spin flipped and on-site modulation of the system. (b) The conditions under which Avila's global theory yields
analytical characterization of the (reduced) 1D quasiperiodic chain.
The nonzero part of derivative of the complexified Lyapunov exponent $\gamma_{\epsilon}$
is quantized to the unique integer: $d\gamma_{\epsilon}/d\epsilon=2\pi\mathbb{Z}$. Then the
original Lyapunov exponent can be obtained by the expression obtained
from the $\epsilon\rightarrow\infty$.
}
\end{figure*}

\subsection{Summary of results }

In this work, we propose the generic spin-$1/2$ QP system, with which we establish a unified theoretical framework for all fundamental phases in Anderson localization physics based on the exact and universal results obtained in this system, and further predict novel new localization physics. The proposed generic QP  framework unifies not only the existing 1D spinful QP models with exact analytical solutions, but also the spinless QP models through an additional Majorana representation. Most importantly, this unified system establishes a versatile platform for constructing new exactly solvable QP models exhibiting rich phase structures, including all pure phases and all coexisting phases with MEs.

The main results are obtained in this unified framework and can be summarized as follows. First, we show with duality transformation and renormalization group method that  pure extended, localized, or critical phases are obtained when  system preserves a chiral(-like) symmetry. Accordingly, the emergence of coexisting phases with MEs can be obtained by breaking the chiral(-like) symmetry. 
Second, we uncover a generalized universal mechanism for critical states in the spinful quasiperiodic system, which arise from generalized incommensurate zeros (GIZs) in matrix elements. This mechanism opens up a much broader way of rigorously realizing and characterizing the exotic critical states. Third, we establish the exactly solvable condition, under which the 1D spinful QP system can be transformed into effectively spinless QP models of dressed particles, and then can be exactly solved. 
These results provide a powerful guiding principle to construct exactly solvable models for novel localization physics, particularly the spin-selective QP model and QP optical Raman lattice model encompassing all basic types of MEs and all seven fundamental localization phases, respectively. 

\subsection{Organization of the work }

The remainder of the paper is organized as follows. In Sec.~\ref{sec:Generic-Framework}, we introduce the generic framework for 1D spinful QP chains, including the basic Hamiltonian and the analytical approaches employed in this study, such as dual transformations, renormalization group methods, and Avila's global theory. In Sec.~\ref{sec:Universal-results}, we present the universal results for the spinful QP chains, as organized into three theorems shown here. These results serve as guiding principles for constructing multiple exactly solvable models hosting distinct classes of localization physics and anomalous MEs, including spin-selective QP model and QP optical Raman lattice model in Sec.~\ref{sec:Exactly-solvable-models}. In Sec.~\ref{sec:Scheme-Exp}, we propose an experimental scheme for the realization based on a generic quasiperiodic optical Raman lattice. Finally, we conclude in Sec.~\ref{sec:Conclusion}, with additional details and proofs provided in Appendix.

\section{Generic Framework}
\label{sec:Generic-Framework}

\subsection{The spinful quasiperiodic model}

We begin with the generic one-dimensional (1D) spin-$1/2$ quasiperiodic
system, whose Hamiltonian is given by \begin{equation}
H=\sum_{j,s,s'}\big(c_{j+1,s}^{\dagger}\Pi_{j}^{s,s'}c_{j,s'}+\mathrm{h.c.}\big)+\sum_{j,s,s'}c_{j,s}^{\dagger}M_{j}^{s,s'}c_{j,s'},\label{eq:UniHam}
\end{equation}
where $c_{j,s}^{\dagger}(c_{j,s})$ creates (annihilates) a particle
on site $j$ with (pseudo)spin $s=\uparrow,\downarrow$ that
can denote real spin, sublattices {or particle-hole}. This model includes two key ingredients, as illustrated in Fig.~\ref{fig:Fig1}(a). First, the hopping coupling matrix $\Pi_{j}=\sum_{l}p_{j}^{l}\sigma_{l}$, as written in terms of Pauli matrices $\sigma_{l}$ with $l=\{0,x,y,z\}$, represents the spin-independent (for $l=0$) and
spin-dependent (for $l=x,y,z$) hopping terms. The latter includes both spin-conserved ($l=z$)
and spin-flip ($l=x,y$) terms, corresponding to the spin-orbit coupling in periodic systems. Second, the onsite matrix $M_{j}=\sum_{l}m_{j}^{l}\sigma_{l}$ also involves the spin-independent chemical
potential and the Zeeman coupling terms. The coefficients
$p_{j}^{l}=t^{l}+\mu^{l}V_{j}^{\mathrm{od}}$ and $m_{j}^{l}=\lambda^{l}+V^{l}V_{j}^{\mathrm{d}}$, where $(t^l,\lambda^l,\mu^{l},V^{l})$ are constants, the off-diagonal QP modulation $V_{j}^{\mathrm{od}}=\cos[2\pi\alpha(j+1/2)+k_{y}]$ and the diagonal QP modulation $V_{j}^{\mathrm{d}}=\cos(2\pi\alpha j+k_{y})$. Here $\alpha=\lim_{n\rightarrow\infty}(F_{n-1}/F_{n})=(\sqrt{5}-1)/2$ is an irrational number approached by the Fibonacci
sequence $F_{n}$,
and $k_{y}$ is a constant phase factor. 
In this work, we take
that the QP term has a single frequency. To facilitate our discussions, we define the combinations of Pauli matrices $\sigma_{l}$
as $\sigma_{\pm}=(\sigma_x\pm+i\sigma_y)/2$ and $\Lambda_{\pm}=(\sigma_z\pm\sigma_0)/2$.

This spinful QP system unifies various QP models, including the existing 1D spinful models with analytical solutions, and spinless models by taking the Majorana representation (see the details in Appendix~\ref{app:Models} and Appendix~\ref{app:Mosaic}). More importantly, with this unified framework we shall develop a series of new and exactly solvable models which host various novel quantum phases with or without mobility edges. Below, we outline the three main theoretical approaches employed in this study.

\subsection{Dual transformation}

The dual transformation is a non-local operation similar to the Fourier transformation, and maps the system between real space and dual momentum space. It is defined as $c_{j,s}=(1/\sqrt{L})\sum_{n}e^{-i2\pi n\alpha j}c_{n,s}$.
The extended, localized, and critical states behave differently under this transformation.
For an extended state, the wave function is delocalized in real space, but localized in dual momentum space, whereas the localized state is localized in real space, but delocalized in dual momentum space. These opposing features are exchanged under the dual transformation. In contrast, critical states exhibit behavior distinct from the other two; they remain delocalized in both real and dual spaces. {Under the dual transformation, critical states are mapped to critical states. In particular, for the self-dual cases, they are invariant under the transformation~\cite{footnote_dual}.}
For example, when the system has a QP hopping term $(V_{j}^{\mathrm{od}}a_{j}^{\dagger}a_{j+1}+\mathrm{h.c.})$, which dominates over other hopping couplings and is invariant under the dual transformation due to the presence of incommensurately distributed zeros (IDZs) in the QP hopping modulation $V_j^{\rm od}$, the critical states may emerge and exhibit self-dual for having incommensurate nodes corresponding to the IDZs~\cite{zhou2023c,liu2024g}. These incommensurate nodes in the delocalized wave functions remain unchanged under the dual transformation, driving the delocalized states into critical states. An intuitive picture suggests that IDZs partition the 1D system into multiple segments, where the wave function reorganizes, exhibiting a self-similar pattern and multifractality of the critical states.

The distinct behaviors of the three types of states can be mapped onto a 2D system to get an intuitive physical interpretation, where $x$- and $y$-axes correspond to the real and dual spaces, and the dual transformation exchanges the two axes. In this mapping, the extended state is delocalized along $x$-direction and localized along $y$-direction, and the opposite is true for localized states. The critical state is delocalized in both directions. Thus, the swapping of the two axes transforms the localized and extended states, while leaving the critical states unchanged. We use 1D spinless 1D extended Aubry-Andr{\'e} (EAA) model~\cite{hatsugai1990,han1994,liu2015} as an example, which is characterized by the nearest-neighbor hopping $t_j=t+\mu V_j^{\rm od}$ and on-site potential $V_j=V_0 V_j^{\rm d}$ (see details in Appendix~\ref{app:Models}). By interpreting the phase shift $k_{y}$ as the quasi-momentum
in the $y$-direction and performing a Fourier transformation, the 1D spinless
EAA model maps to a 2D Hofstadter model that describes a particle
moving in a magnetic field \citep{kraus2012,borgnia2023}. In this mapping, the uniform hopping
terms correspond to hopping in the $x$-direction,  while the on-site potential
$V_{j}^{\mathrm{d}}a_{j}^{\dagger}a_{j}$ corresponds to the hopping
in the $y$-direction with a magnetic flux $\exp(i2\pi\alpha x)a_{x,y}^{\dagger}a_{x,y+1} + {\rm h.c.}$. Similarly, the QP
hopping term $(V_{j}^{\mathrm{od}}a_{j}^{\dagger}a_{j+1}+\mathrm{h.c.})$
represents diagonal hopping in the 2D plane under the magnetic flux.
The relative hopping amplitude shapes the cyclotron orbits while maintaining
the flux within each orbit. When the hopping in the $x$- or $y$- direction dominates, the extended or localized states emerge. In contrast, when the hopping in diagonal direction dominates, hence having equal weight in both directions, the critical states are resulted and exhibit delocalized feature in both axes.

\subsection{Renormalization group}

The renormalization group (RG) approach \citep{suslov1982,thouless1983c,ostlund1984,niu1986,niu1990,szabo2018,jagannathan2021,goncalves2022a,goncalves2023a} investigates the relevance of renormalized coefficients by iterating the rational approximations of the QP parameter. {In particular, Ref.~\cite{goncalves2023a} introduced a new RG framework formulated directly in terms of renormalized coefficients, rather than conventional scaling of wave functions, and this new approach will be adopted in the present work.} Specifically, one takes a rational sequence $\alpha^{(n)}=p_n/q_n$ to define a sequence of periodic Hamiltonians $H^{(n)}$, which approach the quasiperiodic regime only at the irrational limit $\alpha=\alpha^{(\infty)}$.
The RG flow to $n\rightarrow\infty$ determines the properties of the states at quasiperiodic regime. With the rational sequence, the energy dispersion of $H^{(n)}$ is a periodic function of quasi-momenta $\kappa_{x}$ and $\kappa_{y}$ (in the above 2D mapped space). To systematically investigate how the relevant coefficients flow as the system size increases, we compute the characteristic determinant $P^{(n)}=\det|{\cal H}^{(n)}-E|$
at a target energy $E$ for a given size $L=q_{n}$, which yields
\begin{align}
P^{(n)}(E;\kappa_{x},\kappa_{y}) & =t_{R}^{(n)}\cos(\kappa_{x}+\kappa_{x}^{0})+V_{R}^{(n)}\cos(\kappa_{y}+\kappa_{y}^{0})\nonumber \\
 & +\mu_{R}^{(n)}\cos(\kappa_{x}+\tilde{\kappa}_{x}^{0})\cos(\kappa_{y}+\tilde{\kappa}_{y}^{0})\nonumber \\
 & +\epsilon_{R}^{(n)}(E,\kappa_x,\kappa_y)+T_{R}^{(n)}(E),\label{eq:RGdispersion}
\end{align}
where $\epsilon_{R}^{(n)}$ represents irrelevant higher harmonic terms. The asymptotic nature of eigenstates at $n\rightarrow\infty$ is determined by comparing the relative magnitudes of renormalized hopping coefficients. For extended states, the dominant hopping is along $x$ axis of the 2D mapped system, yielding $\big|\mu_{R}^{(n)}/t_{R}^{(n)}\big|,\big|V_{R}^{(n)}/t_{R}^{(n)}\big|\rightarrow0$. For localized states, the dominant hopping is along $y$ direction, leading to $\big|t_{R}^{(n)}/V_{R}^{(n)}\big|,\big|\mu_{R}^{(n)}/V_{R}^{(n)}\big|\rightarrow0$. For critical states, all three hopping coefficients are equally relevant, such that $\big|\mu_{R}^{(n)}/V_{R}^{(n)}\big|,\big|\mu_{R}^{(n)}/t_{R}^{(n)}\big|\geq1$. With the RG flow we can determine the transitions between extend, localized and
critical phases, and the mobility edges as the energy dependent transition points.

\subsection{Avila's global theory}

When the Hamiltonian in Eq.~\eqref{eq:UniHam} can be reduced to an effective spinless QP chain of dressed particles, the system may be analytically solved using Avila's global theory~\cite{avila2015}. The global theory enables  the analytical calculation of the Lyapunov exponent (LE), which is the inverse of the localization length.
The method provides a rigorous characterization of extended, localized, and critical states. Specifically, the non-negative LE $\gamma(E)$, quantifies the localization properties of eigenstates of energy $E$. If $\gamma(E)>0$, the state has a finite localization length $\xi(E)=\gamma^{-1}(E)$. For $\gamma(E)=0$, the state is delocalized, corresponding to either an extended or a critical state. Critical states arise when the transfer matrix $T_{j}$ of the delocalized states  becomes singular~\cite{simon1989,jitomirskaya2012}. This singularity can emerge from a universal physical mechanism that the hopping couplings have IDZs~\cite{zhou2023c}, or the QP potential has incommensurately distributed divergence points, which effectively divides the system into infinite segments, acting as IDZs.

Furthermore, Avila's global theory can also be employed to derive the analytical expression for the correlation length of extended states. This is achieved by performing the dual transformation of the system into dual space. Then one can study the localization length of localized states in the dual space, which is equivalent to the correlation length of the corresponding extended states in the real space.

Now we outline the procedure for obtaining the LE using Avila's global theory. For the generic 1D spinful QP Hamiltonian in Eq.~\eqref{eq:UniHam}, we consider the 4-by-4 transfer matrix $T_j$ given by $(\begin{array}{cccc}
	u_{j+1\uparrow}, & u_{j+1\downarrow}, & u_{j\uparrow}, & u_{j\downarrow}\end{array})^{\intercal}=T_{j}(\begin{array}{cccc}
	u_{j\uparrow}, & u_{j\downarrow}, & u_{j-1\uparrow}, & u_{j-1\downarrow}\end{array})^{\intercal} $,
where $u_{j\sigma}$ is the wave function of eigenstate at site $j$ with spin $\sigma $. In general, the 4-by-4 transfer matrix does not guarantee exact solvability. Exact solvability arises when the transfer matrix is reduced to a block-diagonal 2-by-2 form. As we shall show in Sec.~\ref{subsec:theoremIII}, the present spinful system can be reduced to effective spinless QP model of dressed particles when certain local constraint is introduced. Consequently, the effective transfer matrix becomes a 2-by-2 matrix. In this regime, the LE $\gamma_{\epsilon}(E)$ is computed as
\begin{equation}
	\gamma_{\epsilon}(E)=\lim_{n\rightarrow\infty}\frac{1}{2\pi n}\int\ln||{\cal T}_{n,1}(\theta+i\epsilon)||d\theta.
\end{equation}
Here, $E$ is the energy of the corresponding eigenstate, $||A||$ is the norm of the matrix A, i.e. the square root of
the largest eigenvalue of $A^{\dagger}A$, ${\cal T}_{n,1}=\prod_{j=1}^{n}T_{j}$. The $\epsilon$ is the imaginary
part of complexified quasiperiodic phase shift $\theta+i\epsilon$. The LE is generally challenging to compute analytically, even for the simple 2-by-2 transfer matrix $T_{j}$. Nevertheless, a key result from Avila's
global theory is that with the complexification one can analytically compute the LE for all eigenstates, as long as 
the LE $\gamma_{\epsilon}(E)$ is convex, continuous and piecewise
linear with quantized right-derivative, given by
\begin{equation}
\lim_{\epsilon\rightarrow0^{+}}\frac{1}{2\pi\epsilon}[\gamma_{\epsilon}(E)-\gamma_{0}(E)]=\mathbb{Z}.
\end{equation}
In this case, the nonzero part of derivative of LE is quantized
to an integer for all range of $\epsilon$: $d\gamma_{\epsilon}/d\epsilon=2\pi\mathbb{Z}$. This condition also implies that there is no turning point before $\gamma_{\epsilon}$ intersects the $x-$ or $y-$axis, as illustrated in
Fig.~\ref{fig:Fig1}(b). This has an important consequence that the analytic expression for the LE can be determined in the following way: one first computes the complexified LE $\gamma_{\epsilon}$ at the limit $\epsilon \to \infty$, which can be obtained straightforwardly. Then, using the properties that $\gamma_{\epsilon}(E)$ is convex, continuous, and piecewise linear with a quantized right-derivative, the analytic expression of the LE determined at $\epsilon \to \infty$ can be extended to all values of $\epsilon$. Finally, the exact LE is obtained by taking back $\gamma_{\epsilon=0}$, which represents the original physical Lyapunov exponent that characterizes the properties of all eigenstates.

\begin{figure*}[!t]
	\includegraphics{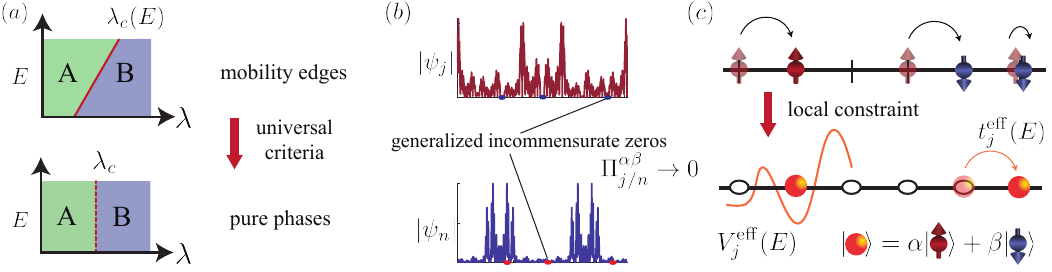}\caption{\label{fig:Fig2}
	Three universal results for the spinful quasiperiodic chains. (a) Criteria for the system to exhibit pure phases without mobility edges. Here, $\lambda$ is the tuning parameter and $E$ is the energy. States A and B refer to extended, localized, or critical states. The energy-dependent transition points $\lambda_c(E)$ become energy-independent, denoted as $\lambda_c$, indicates transitions from a regime with mobility edges (MEs) to a pure phase without MEs. (b) Mechanism for the emergence of critical states in the 1D spinful quasiperiodic (QP) system, generated by dual-invariant generalized incommensurate zeros (GIZs) in matrix elements, marked as the blue and red circles in real and dual spaces. (c) Exact solvability of QP systems from local constraint. The local constraint that reduces the generic 1D spinful QP chain into a 1D spinless QP chain of dressed particles (represented by orange spheres), with energy-dependent or energy-independent nearest-neighbor hopping coefficients $t_{j}^{\mathrm{eff}}(E)$  and on-site potentials $V_{j}^{\mathrm{eff}}(E)$, rendering the system exactly solvable through Avila's global theory.}
\end{figure*}

\section{Universal results for the quasiperiodic spinful chains}
\label{sec:Universal-results}
In this section, with the above theoretical approaches, we show three universal physical results (\emph{theorems}) for the spinful
quasiperiodic systems, which provide the cornerstone to establish a unified framework of all the fundamental localization phases, as summarized in Fig.~\ref{fig:Fig2}. First, we prove a generic symmetry condition, under which the MEs disappear in the system, namely, the pure phases are obtained. Second, we uncover a new fundamental mechanism that critical states are protected by the generalized incommensurate zeros in matrix elements. Third, we outline the conditions, under which the Avila's global theory can be applied to obtain the full analytic solutions. 
Below we mainly elaborate the key physics for the proof, with more details being given in Appendix.

\subsection{Criteria for pure phases without MEs \label{subsec:theoremI} }

We firsts show that the system hosts only pure phases without MEs under the following conditions: \emph{The system preserves chiral (or chiral-like) symmetry, with the on-site matrix $M_{j}$ being purely quasiperiodic, and
the hopping coupling matrix $\Pi_{j}$ being either uniform or purely quasiperiodic,} characterized by
\begin{equation}
	\sigma_{y}\Pi_{j}'\sigma_{y}^{\dagger}=-\Pi_{j}', \qquad \sigma_{y}M_{j}'\sigma_{y}^{\dagger}=-M_{j}'.
\end{equation}
We show this criterion by proving that all leading coefficients in Eq.\eqref{eq:RGdispersion} are energy-independent, ensuring that the transitions between different types of states are energy independent, and consequently, the system exhibits pure phases without MEs [Fig.\ref{fig:Fig2}(a)]. For convenience. we transform $\Pi_{j}'$ and $M_{j}'$ into the local bases such that the on-site coupling matrix is diagonal, given explicitly as
\begin{equation}
	\Pi_{j}'=p_{j}^{\perp}\sigma_{x}+p_{j}^{z}\sigma_{z}, \qquad M_{j}'=m_{j}^{z}\sigma_{z},
\end{equation}
The proof follows from a careful power counting analysis of the energy dependence in characteristic polynomial $P(E;\kappa_x,\kappa_y)$ in Eq.~\eqref{eq:RGdispersion}.
We consider	 an odd system size $L$ without loss of generality. Due to chiral(-like) symmetry, eigenenergies appear in pairs $(-E,E)$, and therefore $P(E)$ is an even function of energy $E$. In the power counting analysis, the energy $E$ explicitly contributes a power of $E$, while $p_{j}^{\perp}$ and $p_{j}^{z}$ contribute dispersions of $e^{ik_{x}}$, and $m_j$ contributes dispersions of $e^{ik_{y}}$ (with the mapping to 2D system). Here $\kappa_x = Lk_x$ and $\kappa_y = Lk_y$. We utilize the periodic structure to simplify the power counting.
Thus, the characteristic polynomial can be expressed as the determinant of a $2L\times2L$ matrix $|A(E)|$ (details in Appendix~\ref{app:ProofI}), with elements
\begin{equation}
A_{mj}=e^{i2\pi\alpha jm}\Big[m_{j}\sigma_{0}+a_{k_{x},k_{y}}^{11}\Lambda_{+}+a_{k_{x},k_{y}}^{22}\Lambda_{-}-E\sigma_{x}\Big],
\end{equation}
where
$a_{k_{x},k_{y}}^{11(22)}=t_{j-1}^{+(-)}e^{-i(2\pi\alpha m+k_{x})}+c.c.,$ and $t_{j}^{\pm}=p_{j}^{z}\pm ip_{j}^{\perp}$. This ensures the energy and momentum dispersions decouple naturally, eliminating cross terms such as $E e^{ik_{x}}$. Below, we analyze separately the dominant dispersions for extended, localized, and critical states.

For the extended and localized states, characterized by dispersions of the form $\cos n_{x} Lk_{x}$ and $\cos n_{y} Lk_{y}$, respectively, where $n_{x},n_{y} \geq 1 \in \mathbb{Z}$, we show that the dominating coefficients are energy-independent. Specifically, the leading dispersion $\cos Lk_{x}$ and $\cos Lk_{y}$ vanish due to chiral symmetry, as these terms arise from the product of $L$ factors of $e^{ik_{x}}$ and $L$ factors of energy $E$, yielding an odd power of energy $E^{L} \cos Lk_{x}(k_{y})$, which is forbidden. Thus, the dominant terms become $\cos 2Lk_{x}$ and $\cos 2Lk_{y}$, whose coefficients are energy-independent, solely given by diagonal elements of the determinant, leaving no opportunity to involve $E$-dependent terms. Therefore, transitions for extended and localized states do not depend on energy, implying the absence of MEs.

For the critical states, characterized by dispersions of the form $\cos n_{x}Lk_{x}\cos n_{y}Lk_{y}$
with $n_{x},n_{y}\geq1\in\mathbf{\mathbb{Z}}$, we show below that they are energy independent. We consider two distinct scenarios for the chiral symmetry: uniform and purely quasiperiodic hopping matrices $\Pi_{j}$, with $M_{j}$ always purely quasiperiodic (details in Appendix~\ref{app:ProofI}).

In the first scenario, for the uniform $\Pi_{j}$, the leading dispersion $\cos Lk_{x}\cos Lk_{y}$ emerges from products of exactly $L$ factors  of
$p_{j}$ (each contributing $e^{ik_x}$) and $L$ factors of $m_{j}$ (each contributing $e^{ik_y}$). Thus the momentum terms are exhausted, leaving no momentum dependence of the coefficients, namely, being energy-independent.

In the second scenario, where both $\Pi_j$ and $M_j$ are purely QP, the leading dispersions include terms such as $\cos 2Lk_x \cos Lk_y$ 
and $\cos 2Lk_x \cos 2Lk_y$~\cite{footnote_dispersion}, both of which are energy-independent. The dispersion $\cos Lk_x \cos 2Lk_y$  arises from product of
$L$ factors of $p_{j}$ (each contributing $e^{ik_x}e^{ik_y}$) and $L$ factors of $m_{j}$ (each contributing $e^{ik_y}$), containing exactly
$2L$ factors, and thus remains free of energy-dependent terms. Similarly, the dispersion $\cos2Lk_{x}\cos2Lk_{y}$ originates
from product of $2L$ factors of $p_{j}$ (each contributing $e^{ik_x}e^{ik_y}$), and the coefficient is still energy-independent.
Consequently, critical states also exhibit energy-independent transitions, eliminating the presence of MEs.

The proof of the criterion highlights the crucial role of chiral(-like) symmetry in excluding MEs for the present spinful quasiperiodic systems. It therefore provides general guidelines for realizing pure phases without MEs, as well as coexisting phases with MEs. For coexisting phases, one way to realize is to directly break the condition by adding higher QP frequencies to the system. 
Alternatively, one may break the chiral(-like) symmetry of the system by introducing spin-conserved hopping terms or QP chemical potential into the Hamiltonian~\cite{footnote_chiral}.

\subsection{Universal mechanism for the emergence of critical states \label{subsec:theoremII}}

We further show a new fundamental mechanism for the critical states in spinful QP chains: \emph{The spinful quasiperiodic systems can host critical
states if the system possesses generalized incommensurate matrix element zeros ${\cal G}_{\Pi}$},
which are defined by
\begin{equation}
	{\cal G}_{\Pi}=\Big\{ j_{k}\Big|\lim_{L\rightarrow\infty}\Pi_{j_{k}}^{\alpha\beta}=0\Big\},
	\end{equation}
where $\alpha,\beta\in\{1,2\}$ label the matrix element indices. The generalized incommensurate zeros (GIZs) in the above equation refer to the IDZs for any component of the hopping coupling matrix $\Pi_j$ in the thermodynamic limit. This is a nontrivial generalization of the mechanism for critical states in spinless QP chains~\cite{zhou2023c} to the present spinful systems. In spinless systems, the IDZs partition the system into multiple segments, driving the delocalized wave function to reorganize and develop the characteristic self-similar structure of critical states~\cite{zhou2023c,liu2024g}. In the current spinful systems, the dual-invariant GIZs will similarly partition system into incommensurately distributed segments in the internal degree of freedom (spin) and the real space, leading to critical states with multifractal features [Fig.~\ref{fig:Fig2}(b)]. When the QP hopping matrix $\Pi_j$ with GIZs dominates over the on-site QP couplings $M_j$, the critical states emerge. {Crucially, the GIZ mechanism does not require the full coupling matrix $\Pi_{j_k}$ to vanish at the sites $\{j_k\}$, in sharp contrast to the spinless case, where IDZs require the complete vanishing of the scalar hopping amplitude.}

We prove this result by mapping the spinful QP model to a generalized 2D bilayer square lattice system, in the presence of an external magnetic field with an irrational flux threading each unit cell.
Under this mapping, the total Hamiltonian Eq.~\eqref{eq:UniHam} becomes (see Appendix~\ref{app:DualMapping})
\begin{align}
	H =&\sum_{x,y,s,s'}\Big[ t^{ss^{\prime}}a_{x+1,y,s}^{\dagger}a_{x,y,s^{\prime}} +\frac{V^{ss^{\prime}}}{2}e^{-i \varphi x}a_{x,y+1,s}^{\dagger}a_{x,y,s^{\prime}} \nonumber \\
	&+\lambda^{ss^{\prime}}a_{x,y,s}^{\dagger}a_{x,y,s^{\prime}}+ \frac{\mu^{s s^{\prime}}}{2}(e^{-i \varphi x} a_{x+1, y+1, s}^{\dagger} a_{x, y, s^{\prime}}  \nonumber \\
	& +e^{i \varphi x} a_{x+1, y-1, s}^{\dagger} a_{x, y, s^{\prime}})   + {\rm h.c.} \Big], \label{eq:2DspinHam}
\end{align}
where $\varphi=2\pi\alpha$, $s,s'$ denote the layer indices, $t^{s s^{\prime}}$ and $\mu^{s s^{\prime}}$ are the uniform and QP hopping coefficients of elements in $\Pi_{j}$, respectively, and $\lambda^{s s^{\prime}}$ and $V^{s s^{\prime}}$ are the uniform and QP on-site coefficients of elements in $M_{j}$, respectively.
The QP hopping term 
$\mu^{ss'}V_{j}^{\mathrm{od}}c_{j+1s}^{\dagger}c_{js'}+\mathrm{h.c.}$ with GIZs in the Hamiltonian Eq.~\eqref{eq:UniHam} characterizes the intra-layer (for $s=s'$) or inter-layer (for $s\neq s'$) hopping coupling in the diagonal direction $(x+1,y\pm1)\leftrightarrow(x,y)$. In contrast, the uniform hopping matrix elements and onsite terms are mapped to the corresponding couplings along $x$ and $y$ directions, respectively.
When the QP hopping dominates over other terms, e.g. when $\mu^{ss'}\gg t^{ss'}, \lambda^{ss'}, V^{ss'}$, the cyclotron motion must be extended in the diagonal direction, leading to delocalization in both $x$ and $y$ directions (the real space and dual space). The dual transformation  is equivalent to a gauge transformation in 2D bilayer (see Appendix~\ref{app:DualMapping} for details)
\begin{equation}
	a^{\dagger}_{x,y,s}\rightarrow e^{i\varphi xy}a^{\dagger}_{x,y,s}, \ \vec{A}=(0,-\varphi x) \rightarrow \vec{A}=(\varphi y,0),
\end{equation}
which clearly preserves the delocalization of cyclotron orbits in the two directions, rendering the invariance under the dual transformation. Thus these states exhibit self-duality, as the key signature of critical states.

In addition to the GIZs, we present a secondary theorem for rigorous realizing critical states in spinful QP chains by connecting the spinful system to the spinless QP models. Specifically, when the on-site matrix $M_j$ has IDZs in the matrix component which is shared by the hopping term $\Pi_j$, one can reduce the spinful QP chain to an effective 1D spinless model, with the IDZs in $M_j$ being transferred into either IDZs in the hopping coefficients or into incommensurately distributed divergence points in the on-site potentials. We therefore reduce to the mechanism in the spinless system for generating the critical states. For instance, when $\Pi_j \sim t \sigma_+$ is uniform, critical states emerge if $M_j$ host IDZs in its $\sigma_x$ or $\sigma_y$ coefficients. We shall show this principle through the quasiperiodic mosaic models and the new exactly solvable models proposed in the next subsection.

The GIZs ${\cal G}_{\Pi}$ and the IDZs in the matrix component of $M_j$ shared in $\Pi_j$ serve
as two powerful guiding principles to rigorously realize the critical states. These two principles can also be proved in an intuitive way based on the semiclassical approximation (see Supplement~\cite{supplement}), which illuminates the origin of the critical states. Together with the criteria for pure phases, these principles provide insights into construction of spinful QP models that host all the fundamental localization phases.

\subsection{Exact solvability of QP systems from local constraint \label{subsec:theoremIII}}

We now present the third universal result regarding {a sufficient condition for} exact solvability of spinful
QP system: \emph{A spinful quasiperiodic system can have exactly solvable points when the hopping coupling matrix
$\Pi_{j}$, or its dual counterpart $\Pi_{n}$, becomes degenerate, which corresponds to }
\begin{equation}
\det|\Pi_{j}|=0 \quad {\rm or} \quad  \det|\Pi_{n}|=0, 
\label{eq:TheoremIII}
\end{equation}
\emph{together with $p_{j}^{s}$ and off-diagonal terms of $M_{j}$ being either constant or purely quasiperiodic}.
This condition represents a local constraint,
where the spinful QP systems can be reduced to
an effectively 1D spinless QP chain for the \emph{dressed particle} with only
nearest-neighbor hopping, as illustrated in Fig.~\ref{fig:Fig2}(c). In this case, Avila's global theory
can be applied to analytically characterize the localization properties of the dressed particles.

To elaborate the physics of this result, we consider the generic eigen-equation for the current spinful QP system that $H\lvert \Psi \rangle =E \lvert \Psi \rangle$, with $\lvert\Psi\rangle=\sum_{j=1}^L (c^\dagger_{j,\uparrow }u_{j,\uparrow }+c^\dagger_{j,\downarrow }u_{j,\downarrow }) \lvert\mathrm{vac}\rangle$,
where $\vec{u}_j = (u_{j\uparrow}, u_{j\downarrow})^{\intercal}$ denotes the spinor for spin-1/2 particles. Under the constraint described in Eq.~\eqref{eq:TheoremIII}, the hopping coupling matrix takes one of the following fundamental forms: for spin-conserved processes, $\tilde{\Pi}_{j} \sim (\sigma_0 \pm \sigma_z)/2 = \Lambda_{\pm}$; for spin-flipped processes, $\tilde{\Pi}_{j} \sim (\sigma_x \pm i\sigma_y)/2 = \sigma_{\pm}$; or a combination of both processes. 
Here, $\tilde{\Pi}_{j}$ represents the hopping coupling matrix after a local unitary transformation, and the transformed spinor is denoted by $\vec{v}_j$. Such constraint reduces the system
to an effective 1D spinless QP chain with nearest-neighbor hopping
\begin{equation}
	H_{\rm eff} = \sum_{j} t_{j}^{\mathrm{eff}} d_{j}^{\dagger} d_{j+1} + \mathrm{h.c.} + \sum_{j} V_{j}^{\mathrm{eff}} d_{j}^{\dagger} d_{j}, \label{eq:Heff}
\end{equation}
For the spin-conserved constraint, $\tilde{\Pi}_{j}\sim t_j\Lambda_{\pm }$, the spin-up and spin-down particles can dress each other via
\begin{equation}
	v_{j,\downarrow }=\frac{M_j^{21}}{E-M_j^{22}}v_{j,\uparrow }.
\end{equation}
Then the effective coefficients of Eq.~\eqref{eq:Heff} are
\begin{equation}
t_{j}^{\rm eff}=t_j,\quad V_{j}^{\mathrm{eff}}=M_{j}^{11}+\frac{M_{j}^{12}M_{j}^{21}}{E-M_{j}^{22}}.\label{eq:Veff}
\end{equation}
Such on-site potential can serves as effective IDZs, which generates critical states in the system. By tailoring the on-site matrix, one can engineer an effective unbounded potential for the dressed particle, even when all couplings in the system are finite. This arises from resonant coupling between the system's energy $E$ and the on-site modulation of the single species $M_j^{22}$, resulting in a divergent effective on-site potential. The divergence in the potential effectively partitions the 1D system into multiple sub-chains, manifesting IDZs in the hopping, and driving the delocalized eigenstates into critical states. Notably, this resonant coupling mechanism is generic, not limited to the exactly solvable regime, and can be extended to systems beyond the exactly solvable conditions.

For the spin-flipped constraint, $\tilde{\Pi}_j\sim t_j\sigma_{\pm }$, similarly, in the transformed basis
\begin{equation}
	v_{j,\downarrow }=\frac{t_j v_{j+1,\uparrow } + M_j^{21}v_{j,\uparrow }}{E-M_j^{22}},
\end{equation}
the effective coefficients of Eq.~\eqref{eq:Heff} are
\begin{equation}
t_{j}^{\mathrm{eff}}=\frac{t_jM_{j}^{12}}{E-M_{j}^{22}},\ V_{j}^{\mathrm{eff}}=M_{j}^{11}+\frac{t_{j-1}^{2}}{E-M_{j-1}^{22}}+\frac{M_{j}^{12}M_{j}^{21}}{E-M_{j}^{22}}.\label{eq:TVeff}
\end{equation}
The effective hopping and on-site potentials in Eqs.~\eqref{eq:Veff} and \eqref{eq:TVeff} highlight the requirement for $p_j^s$ and the off-diagonal elements of $M_j$ to be either constant or purely quasiperiodic in order to preserve exact solvability, as outlined in Theorem III (Sec.~\ref{subsec:theoremIII}). The effective eigen-equation for the dressed particles involves processes such as $M_j^{12} M_j^{21}$, $t_j M_j^{12}$, and $t_{j-1}^2$. These terms maintain a single QP frequency modulation when $p_j^s$ and off-diagonal elements of $M_j$ are either constant or purely quasiperiodic. However, when these terms are a mixture of constant and QP components, such as in the form $[A + B \cos(2\pi \alpha j)]^2$, they lead to mixed frequency modulations, thereby breaking the exact solvability.

The effective hopping coefficient and on-site potential in Eq.~\eqref{eq:TVeff} also explain the guiding principle from theorem II (Sec.~\ref{subsec:theoremII}), where critical states arise if the system possesses IDZs in the shared components of both coupling matrices. Specifically, for the hopping coupling matrix $\Pi_j \sim t \sigma_{+}$, the presence of IDZs in $M_j^{12}$, which couples to $\sigma_x$ and (or) $\sigma_y$, ensures that the effective 1D system exhibits IDZs, hence leading to the emergence of critical states.  This result is exemplified by the absence (or presence) of critical states in type-I (or type-II) QP mosaic models~\cite{wang2020a,zhou2023c}. 
For a unified characterization, we describe the both types of quasiperiodic mosaic models in the spin-$1/2$ QP framework, which can be written as (see more details in Appendix~\ref{app:Mosaic})
\begin{equation}
	H_{\mathrm{M}} = \sum_{j} \lambda(c^{\dagger }_{j+1}\sigma_{-}c_{j}+\mathrm{h.c.})+\sum_{j}c_{j}^{\dagger}V_j^{\mathrm{M}}c_{j}. \label{eq:MosaicUni}
\end{equation}
Here $c_{j}=(c_{j,\uparrow},c_{j,\downarrow})^{\intercal}$ is the
spinor for the annihilation operators, and $\lambda $ is the uniform hopping strength. The onsite matrix $V_j^{\mathrm{M}}$ distinguishes the type-I and type-II mosaic lattices, respectively given by
\begin{align}
	V_j^{\mathrm{M,I}}&=2 V_0 V_j^{\mathrm{d}}\Lambda_{+}+\lambda\sigma_x, \label{eq:MosaicV1} \\
		V_j^{\mathrm{M,II}}&=2tV_j^{\mathrm{d}}(\sigma_0+\sigma_x),\label{eq:MosaicV2}
\end{align}
with $V_0$ and $t$ being the strength of QP modulation of the on-site matrix for type-I and type-II mosaic models, respectively. In the type-I quasiperiodic mosaic model, there are no IDZs in the shared components, resulting in a spectrum composed solely of extended and localized states~\cite{wang2020a}. In contrast, for the type-II quasiperiodic mosaic model, IDZs appear on $\sigma_x$, which introduces energy-dependent quasiperiodic hopping and an unbounded on-site potential, thereby giving rise to rigorously defined critical states within the spectrum~\cite{zhou2023c}.

The universal results offer a powerful guidance to construct exactly solvable models hosting all types of localization physics. The effective hopping and onsite coupling terms [Eq.~\eqref{eq:Veff}-\eqref{eq:TVeff}] provide a framework for designing the microscopic details of these systems by controlling the energy and coupling between the internal degrees of freedom. For example, by designing the effective potential, one can induce or suppress critical states, as well as manipulate the localization length.

\begin{figure*}[t]
	\includegraphics[scale=1.1]{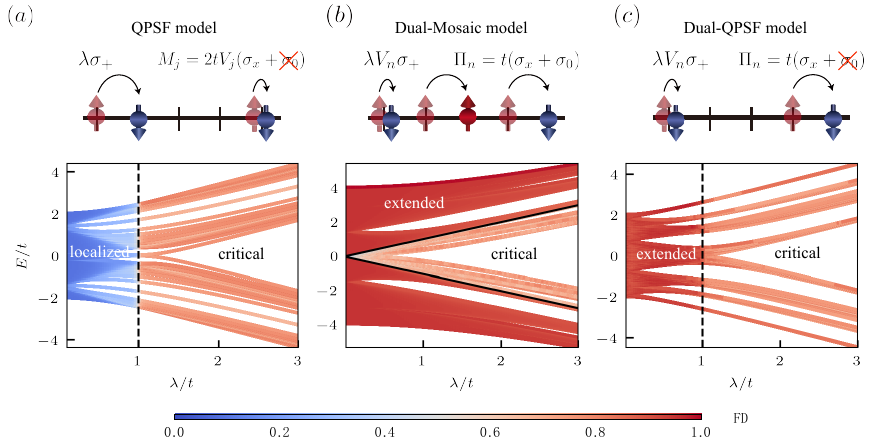}\caption{\label{fig:sshPlusME} New exactly solvable models constructed using universal results and dual transformations. The fractal dimension (FD) versus energy $E$ and hopping strengths $\lambda/t$. (a) The quasiperiodic spin-flipped (QPSF) model, obtained by removing the $\sigma_0$ component of the type-II QP mosaic model, exhibits pure localized and critical phases. (b) The dual counterpart of the type-II QP mosaic model, featuring analytic mobility edges (MEs) that separate extended and critical states. The MEs occur at $E_c = \pm \lambda$ and are marked by solid lines. (c) The dual QPSF model, obtained by removing the spin-independent component in the dual model from (b). It exhibits pure extended and critical phases, with the phase transition points $\lambda = t$ marked by dashed lines. The system size is $L = 2584$.}
\end{figure*}

\subsection{New exactly solvable mosaic models}\label{subsec:examples}

With the key results presented above, we put forward several new exactly solvable models derived from the type-II QP mosaic model~\citep{zhou2023c} in Eq.~\eqref{eq:MosaicUni} and Eq.~\eqref{eq:MosaicV2} by removing the MEs. Through a combination of dual transformation, we can construct new nontrivial models, which demonstrate the applications of the above results, and investigate the phase diagram of the models both analytically and numerically. We numerically solve the
spectrum and employ the fractal dimension (FD) to phenomenologically
characterize the localization properties of the eigenstates $\lvert\Psi\rangle=\sum_{j=1}^{L}u_{j}a_{j}^{\dagger}\lvert\mathrm{vac}\rangle$,
where $\mathrm{FD}=-\lim_{L\rightarrow\infty}\ln\sum_{j=1}^{L}|u_{j}|^{4}/\ln L$,
with $u_{j}$ being the wave function coefficients, $L$ the system
size. In 1D, the FD approaches $1$ for extended states and $0$ for localized
states, while for critical states, $0<\mathrm{FD}<1$. The FD quantifies
the effective dimension experienced by an eigenstate: extended
states uniformly spread across the system, yielding $\mathrm{FD}=1$. Localized
states decay exponentially, effectively perceiving zero dimension,
hence $\mathrm{FD}=0$. Critical states, however, exhibit self-similar
wave function structures that span the entire system, resulting in
multifractality with FD between $0$ and $1$.

We begin with the type-II QP mosaic model~\citep{zhou2023c} in Eq.~\eqref{eq:MosaicUni} and Eq.~\eqref{eq:MosaicV2}, whose Hamiltonian is given by
\begin{equation}
	H_{\mathrm{M-II}} = \sum_{j} \lambda(c^{\dagger }_{j+1}\sigma_{-}c_{j}+\mathrm{h.c.})+2t\sum_{j}V^{\mathrm{d}}_{j}c_{j}^{\dagger}(\sigma_{0}+\sigma_{x})c_{j},\label{eq:mosaicII}
\end{equation}
where $\lambda $ is the uniform hopping strengths and $t$ is the strength of balanced QP potential and exchange coupling. The spectrum comprises critical (localized) states for the energies
satisfy $E<|\lambda|$ ($E>|\lambda|$)~\citep{zhou2023c}.
Next, we construct new exactly solvable models that host pure critical and localized phases by eliminating the MEs, as outlined in theorem I (Sec.~\ref{subsec:theoremI}). This is accomplished by removing the $\sigma_0$-component of the on-site matrix of the type-II QP mosaic model. This results in a QP spin-flipped (QPSF) model with chiral symmetry as
\begin{equation}
	H_{\mathrm{QPSF}} = \sum_{j} \lambda(c^{\dagger }_{j+1}\sigma_{-}c_{j}+\mathrm{h.c.})+2t\sum_{j}V_{j}^\mathrm{d}c_{j}^{\dagger}\sigma_{x}c_{j}.\label{eq:QPSF}
\end{equation}
This model is exactly solvable, as the hopping coupling matrix satisfies $\det |\sigma_{\pm}|=0$. Applying Avila's global theory (see Appendix~\ref{app:ExactSolvable} for details), we find that the system is in the localized
phase when QP spin-flipped process dominates $|t|>|\lambda|$, with the analytic localization length $\xi_l=2/\log|V/t|$ for all eigenstates. In contrast, when $|t|<|\lambda|$, the system is in the critical phase. The numerical results, as shown in Fig.~\ref{fig:sshPlusME}(a), are consistent with the analytical results, with the dashed line marking the phase transition at $\lambda = t$.

The physics can be understood as follows: In the limit $\lambda\rightarrow0$, the QP on-site term localizes the system into dimmers with eigenenergies $E=\pm 2t V_j^{\rm d}$. In this case, the energy difference between adjacent site fluctuates significantly, and turning on the uniform hopping $\lambda$ only hybridize a few dimmers with close energies, giving a localized phase. Conversely, in the limit $t\rightarrow 0$, the uniform spin-flipped hopping creates degenerate dimers with flat band energies $E=\pm \lambda$.  The degeneracy of these dimers is then lifted by the QP on-site term $V_j^{\rm d}$, which hybridizes them drastically, leading to a delocalized phase. The delocalized states are guaranteed to be critical due to the IDZs in the effective hopping
\begin{equation}
	t_j^{\mathrm{eff}}=\lambda V_j^{\mathrm{d}}/{E}.	
\end{equation}

Furthermore, we can develop new models by performing dual transformations
to the two aforementioned models, which exhibit mobility edges
or phase transitions between critical and extended states. Applying the dual transformation to the type-II QP mosaic model in Eq.~\eqref{eq:mosaicII} results in the following dual counterpart
\begin{equation}
	\mathrm{Dual}\big[H_{\mathrm{M-II}}\big] = \sum_{n} t c^{\dagger }_{n+1}(\sigma_{x}+\sigma_{0})c_{n}+\lambda c_{n}^{\dagger} V_{n}\sigma_{+}  c_{n} + \mathrm{h.c.} ,\label{eq:mosaicII-dual}
\end{equation}
with $V_{n}=\lambda\exp(i2\pi\alpha n)$. The model exhibits analytic
MEs at $E_{c}=\pm\lambda$, with extended states for $E>|\lambda|$
and critical states for $E<|\lambda|$, as illustrated in Fig.~\ref{fig:sshPlusME}(b). The absence of the localized orbitals can be understood as follows: under the local rotation $U = \exp(-i\pi \sigma_y / 4)$, the coupling matrices of the dual model become
\begin{equation}
	\tilde{T}_n=2t\Lambda_{+},\ \tilde{M}_n=\lambda \big[\cos(2\pi\alpha n)\sigma_z+\sin(2\pi\alpha n)\sigma_y\big].
\end{equation}
In the limit $t\rightarrow 0$, the system yields two flat bands from the on-site matrix with energies $E = \pm \lambda$. The quantum states of the flat bands can be combined into either localized, extended, or critical states. The extended states and critical states are driven by the uniform hopping and the IDZs in the shared component, respectively.

Similarly, the dual QPSF
model can be obtained by
\begin{equation}
	\mathrm{Dual}\big[H_{\mathrm{QPSF}}\big] = \sum_{n} t c^{\dagger }_{n+1}\sigma_{x}c_{n}+\lambda c_{n}^{\dagger} V_{n}\sigma_{+}  c_{n} + \mathrm{h.c.} ,\label{eq:QPSF-dual}
\end{equation}
with  $V_{n}=\lambda\exp(i2\pi\alpha n)$,
which is equivalent to removing spin-independent component from the previous dual
model [Eq.~\eqref{eq:mosaicII-dual}]. Through the duality transformation, this model exhibits a phase
transition between pure critical and extended phases, with the critical
phase for $|\lambda|>|t|$ and the extended phase for
$|\lambda|<|t|$, with the correlation length given by $\xi_c=2/\log|V/t|$. The numerical results agree with the analytic predictions as shown in Fig.~\ref{fig:sshPlusME}(c). This demonstrates Theorem III in Sec.~\ref{subsec:theoremIII}, which states that the model can be analytically characterized by investigating its dual counterpart.

\section{Exactly solvable models for all fundamental localization phases}
\label{sec:Exactly-solvable-models}

From the above subsection we can see that the universal results provide a powerful guidance to construct new exactly solvable models with nontrivial localization physics. In this section, we present the further in-depth study in developing the highly novel models with exact solutions, which enable a comprehensive and unified characterization of all the basic types of MEs and pure phases in quasiperiodic lattices.
In particular, we propose two classes of exactly solvable models. The first is 1D spin-selective QP (SSQP) lattice model which is shown to host all basic types of MEs, and the second is the QP optical Raman lattice model which hosts all the seven fundamental localization phases. The seven fundamental phases are the three pure phases (extended, critical, and localized), three coexisting phases for any two of the three types of states, and one coexisting phase with all the three types of states. These phases are fundamental, since any generic phase diagram of a QP system can be a certain complex combination of the seven phases.

In the following, we first study the the exactly solvable models for all types of fundamental MEs. Then we show that the QP
optical Raman lattice model can host all the seven fundamental localization phases.

\subsection{The spin-selective QP lattice model for all fundamental MEs\label{subsec:allMEs}}

\begin{figure*}[!t]
	\includegraphics[scale=1.2]{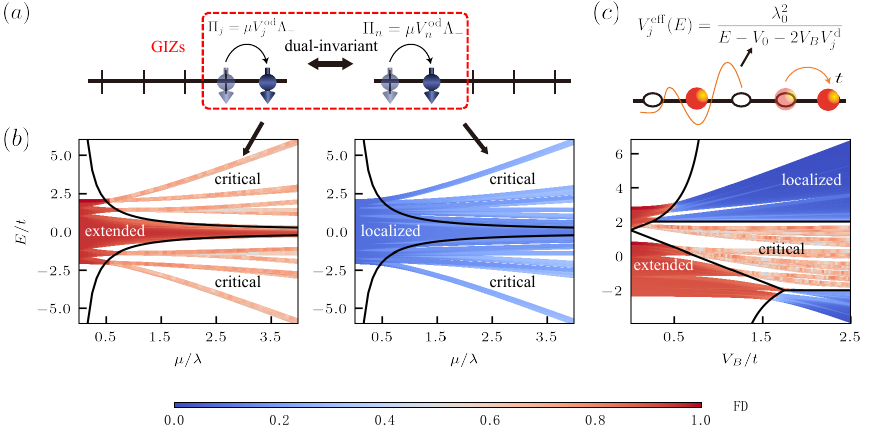}\caption{\label{fig:GIZ}Generalized incommensurate zeros
	generated critical states. (a) The quasiperiodic (QP) spin-conserved hopping in this model gives rise to generalized incommensurate zeros (GIZs), which introduce critical orbitals. (b) The spin-selective QP (SSQP) lattice at the exactly solvable regime. Left panel: The model exhibits analytic mobility edges (MEs) at $E_c = \pm \lambda^2 / \mu$, marked by the solid lines, which separate the extended states ($E < |\lambda^2 / \mu|$) with fractal dimension $\mathrm{FD}\rightarrow 1$, and the rigorous critical states ($E > |\lambda^2 / \mu|$) with $\mathrm{FD}$ approaching a value between 0 and 1. Right panel: The corresponding $\mathrm{FD}$ for the dual model. The MEs $E_c = \pm \lambda^2 / \mu$ (solid lines) now separate the localized states ($E < |\lambda^2 / \mu|$) and the critical states ($E > |\lambda^2 / \mu|$). (c) The SSQP model at another exactly solvable regime that realizes all basic types of MEs. Upper panel: The energy-dependent effective potential resulted from the incommensurately distributed zeros in the shared component, which gives  incommensurately distributed divergent potential. Lower panel: The $\mathrm{FD}$ of the eigenstates shown versus $V_B / t$ and $E/t$, with $\lambda_0 / t = 1$ and $V_0 / t = 1.5$. The MEs are indicated by the solid lines. All systems have a size of $L = 2584$. }
\end{figure*}
The SSQP lattice model for spin-$1/2$ fermions includes several ingredients.
First, the QP hopping or QP onsite potential of the model is considered only for fermions at one spin state, while fermions at another spin state are coupled through the uniform spin-flip transitions. Second, the GIZs are introduced to the hopping matrix $\Pi_j$ or the shared component of $M_j$, leading to the emergence of the critical states.
Third, the spin-independent coupling term is applied to break chiral symmetry, such that MEs emerge. The Hamiltonian is taken in a generic form
\begin{eqnarray}\label{eq:GIZs}
	\begin{split}
		H & =\sum_{j}\Big[c_{j+1}^{\dagger}\Big(t\Lambda_{+}+\mu V_{j}^{\mathrm{od}}\Lambda_{-}+\lambda_{1}\sigma_{x}\Big)c_{j}+\mathrm{h.c.}\Big] \\
 & +\sum_{j}c_{j}^{\dagger}\Big[\big(V_{0}+2V_{B}V_{j}^{\mathrm{d}}\big)\Lambda_{-}+\lambda_{0}\sigma_{x}\Big]c_{j}.
	\end{split}
\end{eqnarray}
We first consider the exactly solvable regime that hosts MEs separating critical and extended states, with its dual counterpart hosting MEs separating critical and localized states. In particular, we take the condition $t=V_0=V_{B}=0$ and $\lambda_{0}=\lambda_{1}=\lambda$. The SSQP model [Eq.~\eqref{eq:GIZs}] has QP hopping only for the spin-down fermions, whereas the spin-up fermions are coupled through uniform spin-flip tunneling and on-site transitions. In this case, the coupling matrices reduce to
\begin{equation}
\Pi_{j}=\mu V_{j}^{\mathrm{od}}\Lambda_{-}+\lambda\sigma_{+},\ M_{j}=\lambda\sigma_{x}.
\end{equation}
Fig.~\ref{fig:GIZ}(b) left subfigure shows the model exhibits MEs separating critical and extended states at $E_c = \pm \lambda^2 / \mu$.
For energies $E < |\lambda^2 / \mu|$, the system exhibits extended states, while for energies $E > |\lambda^2 / \mu|$, critical states are observed. These results can be derived analytically by investigating the dual counterpart, whose coupling matrix is degenerate $|\Pi_{n}| = 0$ (Sec.~\ref{subsec:theoremII}) and is given by
\begin{eqnarray}
	\begin{split}
		& \Pi_n = \mu V_{n}^{\mathrm{od}}\Lambda_{-},\\
		& M_{n}=\lambda\big[\sigma_{x}+\cos(2\pi\alpha n)\sigma_{x}+\sin(2\pi\alpha n)\sigma_{y}\big].
	\end{split}
\end{eqnarray}
By reducing the system to an effective 1D spinless model, the effective hopping and on-site potential are given by
\begin{equation}
t_{n}^{\mathrm{eff}}=\mu V_{n}^{\mathrm{od}},\ V_{n}^{\mathrm{eff}}=\lambda^{2}(2+2V_{n}^{\mathrm{d}})/E. \label{eq:TeffGIZs}
\end{equation}
Neglecting the constant term in the potential, this effective model can be viewed as 1D extended Aubry-Andr{\'e} (EAA) model~\cite{hatsugai1990,han1994,liu2015} without uniform hopping, where the system is in the localized (critical) phase when the QP potential (QP hopping) dominates (see Appendix~\ref{app:Models} for details). Therefore, the system is in the critical phase for $\mu > \lambda^2 / |E|$ and in the localized phase for $\mu < \lambda^2 / |E|$. The numerical results shown in Fig.~\ref{fig:GIZ}(b) right subfigure are consistent with this analytic study. The MEs, $E_c = \pm \lambda^2 / \mu$, separate the critical states for $|E| < \lambda^2 / \mu$ from the localized states for $|E| > \lambda^2 / \mu$. These results can also be obtained by applying Avila's global theory (see Appendix~\ref{app:ExactSolvable} for details), yielding
\begin{equation}
\gamma(E)=\frac{1}{2}\ln\Big| \big|\lambda^{2}/\mu E\big|+\sqrt{(\lambda^{2}/\mu E)^{2}-1}\Big|,  \label{eq:LE_SSQPi}
\end{equation}
which not only provides the MEs, but also gives the localization length $\xi_l = \gamma^{-1}(E)$ for the localized states.

With the above results in the dual space, we can easily obtain the analytic characterization in the original space. Specifically, for $E < |\lambda^2 / \mu|$, the eigenstate with energy $E$ in real space is extended, with a correlation length $\xi_e = \gamma^{-1}(E)$, as localized states are transformed into extended states under the dual transformation. For $E > |\lambda^2 / \mu|$, the eigenstate with energy $E$ remains critical.

We further consider another more exotic exactly solvable regime 
that realizes all three basic types of MEs, namely the MEs separating extended from localized states, extended from critical states, and critical from localized states. This corresponds to $\mu = \lambda_1 = 0$, for which the QP onsite potential is applied solely to the spin-down fermions in the SSQP model. The spin-up fermions have a uniform hopping, and are coupled to spin-down states through the onsite spin-flip transitions. In this regime the coupling matrices simplify to
\begin{equation}
\Pi_{j}=t\Lambda_{+}, \ M_{j}=(V_{0}+2V_{B}V_{j}^{\mathrm{d}})\Lambda_{-}+\lambda_{0}\sigma_{x}. \label{eq:GIZsII}
\end{equation}
The phase diagram in Fig.~\ref{fig:GIZ}(c) reveals all three fundamental MEs resulting from the interplay between IDZs in shared components ($\sigma_z,\sigma_0$) and the broken chiral symmetry.
When $V_B/t$ is sufficiently large, the spectrum exhibits only critical and localized states, with MEs $E_c = \pm 2t$ separating the critical states ($|E| < 2t$) and the localized states ($|E| > 2t$). For intermediate $V_B/t$, extended orbitals also appear in the spectrum.

The underlying physics of the emergence of the three types of MEs can be achieved by reducing the SSQP model in the current regime [Eq.~\eqref{eq:GIZsII}] to an effective 1D spinless model, where the effective energy-dependent on-site potential induced by IDZs in shared components plays a key role
\begin{equation}
	t_{j}^{\mathrm{eff}}=t,   \ V_{j}^{\mathrm{eff}}=\lambda_{0}^{2}/(E-V_{0}-2V_{B}V_{j}^{\mathrm{d}}).\label{eq:TVeffGIZsII}
\end{equation}
When $|E-V_{0}|\leq2V_{B}$, the system exhibits an incommensurately distributed divergent potential, leading to the critical states. This explains why, when $V_B/t$ is sufficiently large, the system exhibits only critical and localized states. In this regime, the analytic LE cab be obtained as (see Appendix~\ref{app:ExactSolvable} for details)
\begin{equation}
\gamma(E)=\ln\Big|\frac{|E/t|+\sqrt{(E/t)^{2}-4}}{2}\Big|,\label{LE1}
\end{equation}
with which the MEs reads $E_c = \pm 2t$.
This result exemplifies a resonant coupling mechanism for critical states: when the onsite coupling is zero ($\lambda_0 = 0$), the two spin states decouple. The
spin-down states produce a set of localized orbitals with energies $E_{j}=V_0+2V_{B}\cos(2\pi\alpha j)$, while the spin-up states produce extended orbitals. Resonantly hybridizing these orbitals leads to  divergent effective on-site potential, generating critical states.

When $|E-V_{0}|>2V_{B}$, the effective potential is finite, giving extended and localized states. The LE can be analytically obtained as (see Appendix~\ref{app:ExactSolvable} for details)
\begin{equation}
	\gamma(E)=\max\Big\{\ln\Big|\frac{|\chi_{E}|+\sqrt{\chi_{E}^{2}-\chi_{B}^{2}}}{|\chi_{0}|+\sqrt{\chi_{0}^{2}-\chi_{B}^{2}}}\Big|,0\Big\},\label{LE2}
\end{equation}
with $\chi_{B}=2V_{B}/t$, $\chi_{E}=EV_{B}/t^{2}$ and $\chi_{0}=(E-V_{0})/t$. A finite $\gamma(E)$ gives the localized states, 
and $\gamma(E)=0$ corresponds to the extended states.

Finally, by combining transition conditions for different effective potential and the corresponding LEs from Eq.~\eqref{LE1} and Eq.~\eqref{LE2}, we determine the analytic MEs that separate extended, localized and critical states, as illustrated in Fig.~\ref{fig:GIZ}(c) and summarized in Table~\ref{Table:threemixed}.
\begin{table}[h!]
	\centering
	\begin{tabular}{|c|c|}
	\hline
	\textbf{Conditions} & \textbf{Phases} \\ \hline
	$|E - V_{0}| > \max\{2V_{B}, |E|V_{B}\}$ & \text{Extended}  \\ \hline
	$2V_{B} > \max\{|E - V_{0}|, |E|V_{B}\}$ & \text{Critical}  \\ \hline
	$|E|V_{B} > \max\{|E - V_{0}|, |2|V_{B}\}$  &  \text{Localized} \\ \hline
	\end{tabular}
	\caption{The criteria for the eigenstates with energies $E$ to belong to one of the three phases: extended, critical, or localized, based on the relationship between the energy $E$, the on-site potential $V_0$, and the quasiperiodic modulation strength $V_B$.}
	\label{Table:threemixed}
\end{table}

\subsection{The model with the seven fundamental localization phases \label{subsec:allPhases}}

\begin{figure*}[t]
	\includegraphics{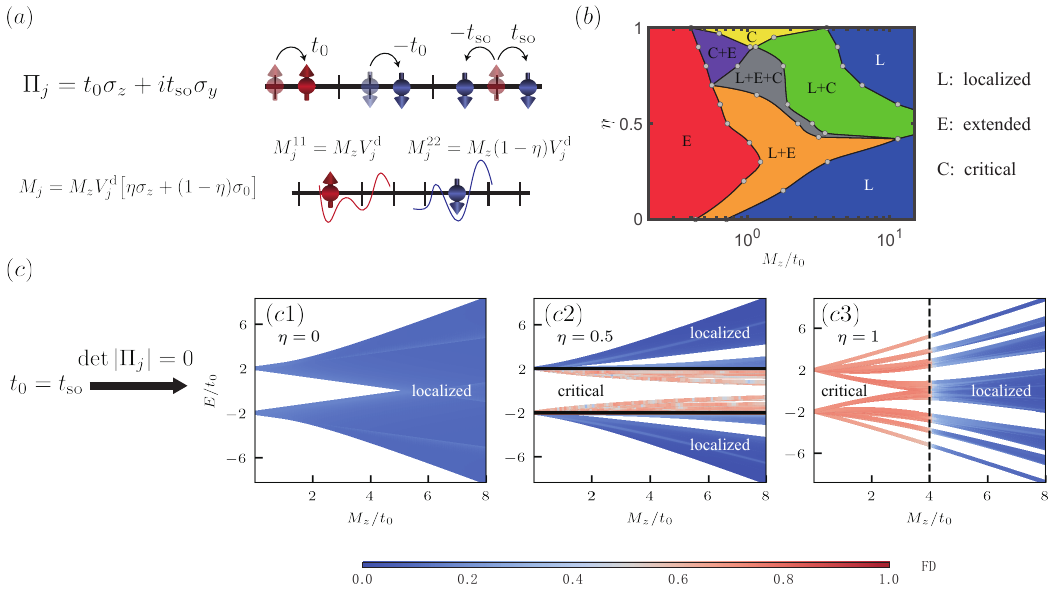}\caption{\label{fig:QPRaman} Seven fundamental phases realized in 1D quasiperiodic (QP) optical Raman lattice. (a) Model illustration: The hopping coupling matrix includes both spin-conserved and spin-flipped hopping terms, while the on-site matrix incorporates both spin-dependent ($\sigma_z$) and spin-independent ($\sigma_0$) QP potential. (b) Phase diagram: The phase diagram, shown as a function of the QP Zeeman potential $M_z$ and the chiral parameter $\eta$, reveals seven distinct phases: three pure phases--extended (E), localized (L), and critical (C)--and four coexistence regions. The latter include three two-coexistence regimes, (L+E), (L+C), and (C+E), where two types of eigenstates coexist at different energies, as well as a three-phase coexistence region (L+E+C). The phase diagram is obtained with the parameter set $t_{\mathrm{so}} = 0.8 t_0$. (c) Exact Solvable Points: (c1) When $\eta = 0$, the entire spectrum is localized. (c2) For $\eta = 0.5$, the spectrum splits into localized and critical states, with the mobility edges marked by solid lines at $E_c = \pm 2t_0$. States with $|E| < 2t_0$ are critical, while those with $|E| > 2t_0$ are localized. (c3) At $\eta = 1$, the system exhibits pure localized and critical phases, with the transition point marked by the dashed line at $M_z = 4t_0$. The system is in the localized (critical) phase when $M_z > 4t_0$ ($M_z < 4t_0$). The system size used in these calculations is $L = 2584$. } 	
\end{figure*}

We now show a highly novel result that the seven fundamental localization phases can be realized in the 1D QP optical Raman lattice model~\citep{wang2020b}, as illustrated in Fig.~\ref{fig:QPRaman}(a). This model is constructed by manipulating the IDZs in the shared components and the chiral symmetry. Specifically, in the presence of chiral symmetry, three distinct pure phases are obtained. Breaking chiral symmetry introduces MEs and four additional coexisting phases. The Hamiltonian reads
\begin{align}
	H & =\sum_{j}\Big[c_{j+1}^{\dagger}(t_{0}\sigma_{z}+it_{\mathrm{so}}\sigma_{y})c_{j}+\mathrm{h.c.}\Big]+\nonumber \\
	 & +M_{z}\sum_{j}c_{j}^{\dagger}\Big[\eta V_{j}^{\mathrm{d}}\sigma_{z}+(1-\eta)V_{j}^{\mathrm{d}}\sigma_{0}\Big]c_{j},\label{eq:1DSOC}
\end{align}
where $t_0$ ($t_{\mathrm{so}}$) denote spin-conserved (spin-flipped) hopping coefficient, $M_z$ is QP potential strength, and $\eta$ is the chiral parameter that controls the ratio of spin-dependent (Zeeman) to spin-independent (chemical) QP potentials, governing the extent of chiral symmetry breaking.

Fig.~\ref{fig:QPRaman}(b) shows the phase diagram versus $M_z$ and $\eta$, which shows the presence of the seven fundamental phases. These results are first obtained by numerically diagonalizing the model in Eq.\eqref{eq:1DSOC} with $t_{\mathrm{so}}=0.8t_0$. We examine two limiting cases: exact chiral symmetry ($\eta = 1$) and complete chiral symmetry breaking ($\eta = 0$), followed by a discussion of the intermediate value of $\eta$. In the case of complete chiral symmetry breaking $\eta=0$, corresponding to a purely spin-independent QP potential, the system exhibits extended and localized phases in the weak and strong quasiperiodic potential regimes, respectively. In this case, the transition energies remain outside the eigenstate spectrum, preserving the pure phases. For the intermediate value of $\eta$, a coexisting phase (L+E) emerges, with MEs separating extended and localized states. This result can be understood as follows: in the absence of spin-flipped hopping ($t_{\mathrm{so}} = 0$), the model in Eq.\eqref{eq:1DSOC} describes two decoupled spin-up and spin-down QP chains with opposite energy spectra,  where each chain hosts either purely localized or extended phases. Turning on $t_{\rm so}$ couples the two chains, opening gaps within each and flattening the dispersion of original band states. This coupling localizes part of states in the moderate potential regime, leading to the formation of MEs. On the other hand, in the chiral symmetry limit $\eta = 1$, which corresponds to purely QP Zeeman potential, distinct phases appear in the weak, moderate, and strong field regimes, corresponding to pure extended, critical, and localized phases, respectively.  These results are consistent with the predictions of Theorem I (Sec.\ref{subsec:theoremI}) and previous studies~\cite{wang2016b,wang2020b}. As indicated by theorem I and II, coexisting phases involving critical states (C+E, L+C, L+E+C) necessitate introducing generalized incommensurate zeros in matrix elements and breaking chiral symmetry. This corresponds to the presence of both spin-dependent and spin-independent QP potentials. Fig.~\ref{fig:QPRaman}(b) indicates that these phases occur for $\eta_c<\eta<1$, with $\eta_c\approx0.43$.

The emergence of the seven fundamental phases can be analytically predicted by combining the universal results and Avila's global theory. By applying Theorem III in Sec.~\ref{subsec:theoremIII}, the system is reduced to a spinless 1D QP chain with nearest-neighbor hopping when $\det|\Pi_j|=0$, which occurs when $|t_0| = |t_{\mathrm{so}}|$. We set $t_{0}=t_{\mathrm{so}}$ without losing generality. The system exhibits chiral symmetry at $\eta=1$, where only pure critical and localized phases exist when $t_{\rm so}=t_{0}$, and all three pure phases are present when $t_{\rm so}\ne t_{0}$. Introducing a spin-independent quasiperiodic potential ($\eta\neq 1$) breaks chiral symmetry, leading to another exactly solvable point at $\eta=1/2$, where MEs emerge within the original critical and localized phases for $t_{\rm so}=t_{0}$, giving rise to the L+C phase. {Away from these exactly solvable limits, the remaining coexisting phases are obtained by continuously deforming the analytically established phase boundaries and are further supported by numerical calculations.} Further, various coexisting phases with combinations of the three types of quantum states emerge for $\eta < 1$ when $t_{\mathrm{so}} \neq t_0$. Therefore, both numerical and analytical results confirm the existence of the seven fundamental localization phases, establishing this system as a novel quantum platform for exploration of localization physics.

In the following, we elaborate the  entire phase diagram, starting from the high symmetry line $t_0=t_{\mathrm{so}}$, where the effective nearest-neighbor hopping is given by
\begin{equation}
	t_{j}^{\mathrm{eff}}=\frac{-2t_{0}\eta\Delta_{j}}{E-(1-\eta\Delta_{j})},
\end{equation}
with $\Delta_j=M_z V_j^{\mathrm{d}}$. The energy-dependent effective on-site potential is given by
\begin{equation}
	V_{j}^{\mathrm{eff}}=\frac{4t_{0}^{2}}{E-(1-\eta)\Delta_{j-1}}+\frac{E(1-\eta)\Delta_{j}-(1-2\eta)\Delta_{j}^{2}}{E-(1-\eta)\Delta_{j}}.
\end{equation}
We first analyze the three limiting cases, $\eta = 0$, $\eta = 1$, and $\eta = 1/2$ while keeping $t_0=t_{\rm so}$, and then extend the discussion to cases where $\eta$ deviates from these limits and $t_0 \neq t_{\mathrm{so}}$, to explain the entire phase diagram.

\subsubsection{Pure spin-independent quasiperiodic potential}

For pure spin-independent limit of QP potential, where $\eta=0$, the system exhibits only extended and localized states, with no critical states since IDZs in the shared component vanish. We first consider the high-symmetry case where $t_0 = t_{\mathrm{so}}$, which leads to a vanishing effective hopping, leaving only the on-site potential
\begin{equation}
	t_j^{\mathrm{eff}} = 0, \ V_{j}^{\mathrm{eff}}=\frac{4t_{0}^{2}}{E-\Delta_{j-1}}+\Delta_j.
\end{equation}
In this case, all eigenstates are localized as shown in Fig.~\ref{fig:QPRaman}(c1). The mechanism behind this can be understood as follows: when $M_z = 0$ (i.e., no QP potential), the Hamiltonian [Eq.~\eqref{eq:1DSOC}] yields two flat bands, and any finite QP onsite potential $M_z$ can fully localize all states. Deviating from the high-symmetry line $t_{0}\neq t_{\mathrm{so}}$ introduces next-nearest-neighbor (NNN) hopping $t_j^{\mathrm{NN}} a_{j}^{\dagger}a_{j+2}+{\rm h.c.}$, with coefficient
\begin{equation}
	t_j^{\mathrm{NN}}=\frac{t_{+}t_{-}}{E-\Delta_{j+1}},
\end{equation}
where $t_{\pm}=t_{0}\pm t_{\mathrm{so}}$. In this regime, nearest-neighbor hopping remains zero $t_j^{\rm eff}=0$, and on-site potential becomes
\begin{equation}
	V_j^{\mathrm{eff}}=\frac{t_{+}^{2}}{E-\Delta_{j-1}}+\frac{t_{-}^{2}}{E-\Delta_{j+1}}+\Delta_{j}.
\end{equation}
The emergence of NNN hopping begins to destabilize localization, allowing for the appearance of delocalized states. Although the system is no longer analytically solvable for $t_0 \neq t_{\rm so}$, the possible phases can still be determined. For sufficiently large (weak) $M_z$, the system always enters localized (extended) phase. On the other hand, for intermediate $M_z$ values, the chiral symmetry is explicitly broken by spin-independent QP modulation $M_z V_j^{\mathrm{d}} \sigma_0$, leading to the emergence of MEs between the extended and localized states. This accounts for extended, localized and L+E phases in Fig.~\ref{fig:QPRaman}(b) at $\eta=0$, which extend to small $\eta$ regime with $\eta<\eta_c\approx0.43$.

\subsubsection{Pure spin-dependent quasiperiodic potential}

For pure spin-dependent QP potential with $\eta=1$, we begin by considering the case $t_0 = t_{\mathrm{so}}$, which yields the following effective hopping and on-site potential
\begin{equation}
	t_j^{\mathrm{eff}}=-2t_0\Delta_j,\  V_j^{\mathrm{eff}}=4t_0^2+\Delta_{j}^2.
\end{equation}
The corresponding Lyapunov exponent (LE) is given by
\begin{equation}
\gamma=\max\left\{\frac{1}{2}\ln\left|\frac{M_{z}}{4t_{0}}\right|,0\right\}.
\end{equation}
When $|M_{z}|>4|t_{0}|$, the QP potential dominates, leading to a localized phase, where all states exhibit a localization length $\xi = \gamma^{-1}$.
When $|M_{z}|<4|t_{0}|$, the LE vanishes ($\gamma = 0$), and the system enters the critical phase.  This behavior is attributed to the IDZs in the hopping coefficients, consistent with the numerical calculations shown in Fig.~\ref{fig:QPRaman}(c3). No extended states are observed in this regime when $\eta = 1$ and $t_0 = t_{\mathrm{so}}$ as expected.

We extend our analysis to the case where $t_0 \neq t_{\mathrm{so}}$,  which introduces modifications to the effective nearest-neighbor hopping and on-site potential due to the imbalance between $t_0$ and $t_{\mathrm{so}}$
\begin{equation}
	t_j^{\mathrm{eff}}={t_{-}\Delta_{j+1}-t_{+}\Delta_{j}},\ V_j^{\mathrm{eff}}={2t_0^2 + 2t_{\mathrm{so}}^2+\Delta_{j}^2},
\end{equation}
and more importantly, such imbalance introduces an effective NNN hopping, with coefficient given by
\begin{equation}
	t_j^{\mathrm{NN}}=t_{+}t_{-}.
\end{equation}
This NNN hopping, induced by the imbalance between $t_0$ and $t_{\mathrm{so}}$, gives rise to extended orbitals within the phase diagram. When the imbalance $|t_0 - t_{\mathrm{so}}|$ is small relative to $M_z$, the critical states, generated by zeros in the hopping terms, remain dressed by the NNN hopping $t_j^{\mathrm{NN}}$, preserving the critical phase. However, as the imbalance increases and $t_j^{\mathrm{NN}}$ dominates, the critical states transition into extended states~\cite{huang2025}.

At $\eta=1$, chiral symmetry removes the MEs, leaving only three pure phases: localized, extended, and critical. The global theory does not strictly apply here, however, since the system has no MEs, the phase boundaries can still be determined by analyzing the typical eigenstates. For example, by considering the zero-energy states $E=0$,  the eigen-equation Eq.~\eqref{eq:1DSOC} for  spin-up (spin-down) states $\psi_j$ ($\varphi_j$)
can be expressed in the form of transfer matrix as $(\psi_{j+1},  \psi_{j})^{\intercal}=A_{j}(\psi_{j},  \psi_{j-1})^{\intercal}$, with $A_j$ the transfer matrix at site $j$.
In the pure spin-dependent limit ($\eta=1$), the delocalized phase
is topological while the localized phase is topologically trivial~\cite{wang2020b}. The phase boundary between the critical and localized phases can thus be determined by the topological transition. If both eigenvalues of the transfer matrix $A=\prod_{j=1}^{L}A_{j}$ are either less than 1 or greater
than 1, the system is topological, and the $\psi$
zero mode is localized at one end of the chain. Assuming $t_{0}>0$ and $t_{\mathrm{so}}>0$, then the two eigenvalues
of $A$ satisfy $|\lambda_{1}\lambda_{2}|<1$ , the topology is dictated by the larger eigenvalue $|\lambda_{2}|$.
By a similar transformation~\cite{degottardi2011a,degottardi2013a} or applying the global theory, we find the localized-to-critical transition point is given by $|M_{z}/2t_+|=1$.
By applying the same reasoning to the transfer matrix in the dual space, the phase boundary between localized and critical phases in the dual space is $|M_{z}/2t_{-}|=1$,
corresponding to the transition between extended and critical phases in the original space. Therefore, for $\eta=1$ and $t_{0}\neq t_{\mathrm{so}}$, the system hosts three distinct pure phases as summarized in Table.~\ref{Table:1DSOC}.
\begin{table}[h!]
	\centering
	\begin{tabular}{|c|c|}
	\hline
	\textbf{Conditions} & \textbf{Phases} \\ \hline
	$M_{z}<2|t_{-}|$ & \text{Extended}  \\ \hline
	$2|t_{-}|<M_{z}<2|t_{+}|$ & \text{Critical}  \\ \hline
	$M_{z}>2|t_{+}|$  &  \text{Localized} \\ \hline
	\end{tabular}
	\caption{Phases of 1D QP optical Raman lattice model at $\eta=1$, as a function of QP Zeeman potential strength $M_z$ and  $t_{\pm}=t_{0}\pm t_{\mathrm{so}}$.}
	\label{Table:1DSOC}
\end{table}

When chiral parameter $\eta$ deviates from $\eta=1$, the chiral symmetry is broken and MEs appear. This leads to the emergence of the C+E phase, which interpolates between the pure critical and extended phases, and the L+C phase, which interpolates between the pure localized and critical phases, as shown in Fig.~\ref{fig:QPRaman}(b). Furthermore, as the chiral parameter $\eta$ is further deviated from 1, the L+E+C phase emerges, interpolating between the C+E and L+E phases.

\subsubsection{Balanced quasiperiodic potential}

For balanced QP potential with $\eta=1/2$, the energy-dependent effective nearest-neighbor hopping and on-site potential are given by
\begin{equation}
	t_{j}^{\mathrm{eff}}=\frac{-2t_{0}\Delta_{j-1}}{2E-\Delta_{j-1}},\  V_{j}^{\mathrm{eff}}=\frac{8t_{0}^{2}}{2E-\Delta_{j-1}}+\frac{E\Delta_{j}}{2E-\Delta_{j}}.
\end{equation}
Applying Avila's theory, the LE for the system is
\begin{equation}
\gamma(E)=\max\Big\{\frac{1}{2}\ln\Big||E/2t_{0}|+\sqrt{E^{2}/4t_{0}^{2}-1}\Big|,0\Big\}.
\end{equation}
The states with energy $|E|>2|t_{0}|$ are localized, with localization length
$\xi(E)=\gamma^{-1}(E)$. The states are critical
when $|E|<2|t_{0}|$, as the $\gamma(E)=0$ and the hopping terms
have IDZs. Consequently, $E=\pm2t_{0}$ are
critical energies separating localized and critical states, indicating the presence of MEs as shown in Fig.~\ref{fig:QPRaman}(c2). Thus, for $\eta=1/2$ and $t_{0}=t_{\mathrm{so}}$, the system is in L+C phase. We extend this analysis to the case where $t_0 \neq t_{\mathrm{so}}$, introducing an effective NNN hopping term given by
\begin{equation}
	t_j^{\mathrm{NN}}=\frac{-2t_{+}t_{-}}{2E-\Delta_{j+1}}.
\end{equation}
Similarly, the imbalance between $t_0$ and $t_{\mathrm{so}}$ generates extended orbitals within the spectrum, while critical states still exist when the NNN hopping $t_j^{\mathrm{NN}}$ is small relative to $M_z$. The combination of extended orbitals and the L+C phase results in pure extended, L+E, and L+E+C phases. Similar phase transitions also occur when $\eta$ deviates from $\eta = 1/2$. This analysis clarifies the origin of the L+C phase at the exactly solvable point $\eta = 1/2$ and $t_0 = t_{\mathrm{so}}$, and demonstrates how the related phases,extended, critical, L+E, and L+E+C phases, emerge near $\eta = 1/2$, as depicted in Fig.~\ref{fig:QPRaman}(b).

\begin{figure*}[!t]
	\includegraphics[scale=0.65]{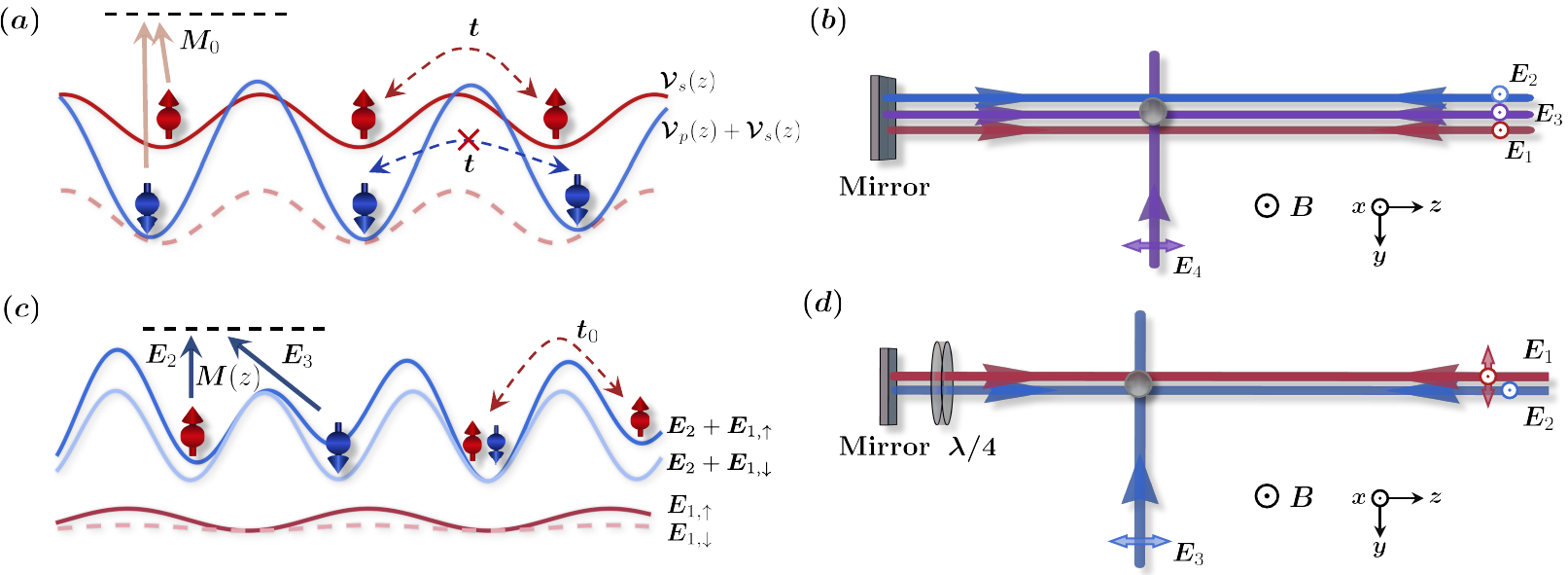}
	\caption{\label{fig:FigSec5_3} Experimental realization schemes. (a)-(b) Implementation of the model supporting all fundamental mobility edges (MEs). (a) The spin-conserved hopping for spin-down atoms is suppressed by a spin-dependent deep primary potential ${\cal V}_p (z)$ (blue solid line), while spin-up atoms undergo hopping in a secondary potential ${\cal V}_s (z)$ (red solid line). The on-site potential for spin-down atoms (red dashed line) is realized by the combination of ${\cal V}_p(z)$ and ${\cal V}_s(z)$ , which are incommensurate with each other. The on-site spin-flipped process is introduced via an overall coupling $M_0$. (b) Schematic of the experimental setup. The spin-dependent potential ${\cal V}_p (z)$ is formed by the combination of ${\bf E}_2$ and ${\bf E}_3$ with the traveling wave ${\bf E}_4$, while the spin-independent secondary potential ${\cal V}_s (z)$ is generated by ${\bf E}_1$, affecting both spin states. (c)-(d) Realization of the quasiperiodic optical Raman lattice model. (c) The spin-conserved hopping $t_0$ is induced by the primary lattice ${\cal V}_p (z)$, and the spin-flipped hopping $t_{\rm so}$ is induced by the Raman potential ${\cal M}(z)$. The quasiperiodic Zeeman potential is realized by the secondary lattice ${\cal V}_s (z)$. (d) Schematic of the setup. A standing wave ${\bf E}_1$, combined with the blue-detuned spin-independent primary lattice generated by ${\bf E}_2$, creates the spin-dependent Zeeman potential. A traveling wave ${\bf E}_3$ generates the Raman potential ${\cal M}(z)$ in combination with ${\bf E}_2$.
	}
\end{figure*}

\subsection{Unified picture for the seven fundamental
localization phases}
We highlight the mapping relation between the present 1D QP optical Raman lattice model in Eq.~\eqref{eq:1DSOC} and the type-II QP mosaic model~\cite{zhou2023c} in Eq.~\eqref{eq:mosaicII}, with slightly modified on-site potential, given by
\begin{equation}
	H_{\mathrm{M}} = \sum_{j} \lambda(c^{\dagger }_{j+1}\sigma_{-}c_{j}+\mathrm{h.c.})+\sum_{j}V^{\mathrm{d}}_{j}c_{j}^{\dagger}(2V_0\sigma_{0}+2t\sigma_{x})c_{j}.\label{eq:mosaicIINew}
\end{equation}
Here, $V_0$ is the strength of the on-site potential. At the high symmetry line $V_0=t$, the model Eq.~\eqref{eq:mosaicIINew} reduces to the origin one in Eq.~\eqref{eq:1DSOC}, with analytic MEs between critical and localized states as discussed in Sec.~\ref{subsec:examples}. This mapping deepens our understanding for the entire phase diagram in Fig.~\ref{fig:QPRaman}(b).

The QP optical lattice model model at $t_{0}=t_\mathrm{so}$ can be mapped to the type-II QP mosaic model [Eq.~\eqref{eq:mosaicIINew}] followed by a local rotation $U=\exp(-i\pi\sigma_y/4)$, with the coefficients between two models are related by the following substitution rule
\begin{eqnarray}
\lambda&\leftrightarrow&2t_{0}, \\
2t &\leftrightarrow&\eta M_{z},\\
2V_{0} &\leftrightarrow&(1-\eta)M_{z}.
\end{eqnarray}
The pure spin-dependent QP potential limit ($\eta=1$) of QP optical Raman lattice model corresponds to the type-II QP mosaic model in the limit $V_0=0$. In this scenario, both models are exactly solvable, exhibiting localized and critical phases. Specifically, the type-II QP mosaic model is in the localized phase when $|t|>|\lambda|$ and in the critical phase when $|t|<|\lambda|$ [Fig.~\ref{fig:sshPlusME}(a)]. This is consistent with the QP optical Raman lattice model, where the localized phase occurs when $M_z>4t_{0}$ and the critical phase when $M_z<4t_{0}$ [Fig.~\ref{fig:QPRaman}(c3)].

Reducing the chiral parameter $\eta$ from 1 to 1/2 corresponds to turning on the on-site potential $V_0$ in the type-II QP mosaic model, transitioning from $V_0=0$ to $V_0=t$. 
{This breaks} the chiral symmetry of both models 
{and introduces} the MEs between localized and critical states, leading to the L+C phase. As the on-site potential increases towards the high-symmetry line $V_0=t$, or when the QP optical Raman lattice model reaches $\eta=1/2$, the system becomes analytically solvable again. The type-II QP mosaic model exhibits analytic MEs $E=\pm\lambda$ separating the localized and critical states~\cite{zhou2023c}, which corresponds to the MEs $E=\pm2t_{0}$ in the QP optical Raman lattice model.

Further introducing the imbalance between $t_{0}$ and $t_{\mathrm{so}}$ leads to the emergence of extended phases and associated MEs.  And this will lead to the long-range hopping terms when reduced the system into the spinless QP chain, introducing extended orbital into the spectrum of the type-II QP mosaic model. Thus, this mapping, combined with the imbalance between $t_{0}$ and $t_{\mathrm{so}}$, accounts for the generation of the seven phases in the phase diagram.

\section{Schemes for experimental realization}
\label{sec:Scheme-Exp}
Finally, we propose experimental schemes for realizing the SSQP on-site model [Eq.~\eqref{eq:GIZsII}] and QP optical Raman lattice model [Eq.~\eqref{eq:1DSOC}]. These schemes utilize ultracold alkali atoms, which are ideal candidates due to their large fine-structure energy splitting and moderate natural linewidth. These properties facilitate  efficient Raman coupling and spin-dependent potentials, even when the laser frequency is far detuned from the $D_1$ and $D_2$ lines~\cite{Soltan2011, Kuzmenko2019, Szulim2022}, making ultracold alkali atoms well-suited for implementing the QP lattice. {We further outline experimentally accessible observables for detecting the distinct phases and the associated mobility edges.}

We first propose the scheme for the SSQP on-site model in Eq.~\eqref{eq:GIZsII} supporting all three basic MEs, as illustrated  in Fig.~\ref{fig:FigSec5_3}(a-b), whose Hamiltonian is
\begin{align} \label{eq:GIZs_Exp}
	H=\left[\frac{p_z^2}{2m}+{\cal V}_s(z)\right]\otimes\sigma_0-{\cal V}_p(z)\Lambda_{-} +M_0\sigma_x,
\end{align}
where the first term represents the kinetic energy.  ${\cal V}_p(z)=V_p\cos(k_pz+\phi_p)$ is a deep spin-dependent primary lattice that freezes the spin-conserved hopping for spin-down atoms. The different terms in Eq.~\eqref{eq:GIZsII} can be realized as follows. The uniform spin-conserved hopping $\Pi_j=t\Lambda_{+}$ for spin-up atoms can be obtained by $t=-\int dzw_{s,\uparrow}^*(z)[p_{z}^{2}/2m+{\cal V}_s(z)]w_{s,\uparrow}(z-a_s)$,
where $w_{s,\uparrow}(z)$ is the Wannier function of the secondary lattice generated by 
potential ${\cal V}_s(z)=V_s\cos(k_sz+\phi_s)$, with $a_s=\pi/k_s$ being the lattice constant. The potential ${\cal V}_s(z)$  is incommensurate with the potential ${\cal V}_p(z)$ for spin-down atoms, leading to the QP potential $M_j=2V_B V_j^{\rm d}\Lambda_{-}$ for spin-down atoms $2V_BV_j^d=\int dz{\cal V}_{s}(z)\abs{w_{s,\downarrow}(z-ja_p)}^{2}=2V_B\cos(2\pi\alpha j+\phi_s)$,
where $\alpha=k_s/k_p$, and $2V_B\equiv \int dzV_s\cos(k_sz)\abs{w_{s,\downarrow}(z)}^{2}$. The Wannier function $w_{s,\downarrow}(z)$ is determined by the primary lattice, with the lattice constant $a_p=\pi/k_p$. Finally, the on-site spin-flipped term $M_j=\lambda_0 \sigma_x$  can be induced via $M_0 \sigma_x$ in Eq.~\eqref{eq:GIZs_Exp}.

The proposed Hamiltonian in Eq.~\eqref{eq:GIZs_Exp} can be realized based on QP optical Raman lattice, as outlined  below. The detailed implementation can be found in Supplement~\cite{supplement}. To facilitate the description, we perform a unitary transformation, such that $\sigma_z\rightarrow \sigma_x$ and $\sigma_x\rightarrow -\sigma_z$. The kinetic term and spin-independent potential ${\cal V}_s(z)$ remain unchanged and can be generated using the standing wave ${\bf E}_1$. The rotated spin-dependent potential ${\cal V}_p(z)(\sigma_0-\sigma_x)/2$  can be realized using a standing waves ${\bf E}_2$ and ${\bf E}_3$ along with a traveling wave ${\bf E}_4$. Finally, the $M_0 \sigma_z$ term arises from the two-photon detuning of the Raman coupling.  After performing the inverse unitary transformation, we reach the Hamiltonian in Eq.~\eqref{eq:GIZs_Exp}. More details can be found in supplement~\cite{supplement}, where  $^{87}{\rm Rb}$ atoms are used as an example.

We then propose the scheme for the model supporting all seven fundamental phases [Eq.~\eqref{eq:1DSOC}] based on the QP optical Raman lattice [Fig.~\ref{fig:FigSec5_3}(c,d)], which has been successfully employed for the realization of synthetic gauge fields and the exploration of topological phases with ultracold atoms~\cite{zhang2018,liu2014,wu2016,wang2018,sun2018,song2019,lu2020,wang2021,zhang2024e}. The Hamiltonian is given by
\begin{eqnarray}
		\begin{split}
			H&=\bigg[\frac{p_{z}^{2}}{2m}+{\cal V}_{p}(z)+(1-\eta){\cal V}_{s}(z)\bigg]\otimes\sigma_{0} \\
		&+{\cal M}(z)\sigma_{x}+\eta{\cal V}_{s}(z)\sigma_{z}.
		\end{split}
\end{eqnarray}
The different terms in Eq.~\eqref{eq:1DSOC} can be realized accordingly. The uniform hopping coupling $\Pi_j=t_0\sigma_z+it_{\rm so}\sigma_y$ arises from the tunneling of the Wannier function of the primary lattice, where $t_0=-\int dzw_s^*(z)[p_{z}^{2}/2m+{\cal V}_{p}(z)]w_s(z-a_0/2)$,
with $p_{z}^{2}/2m$ being the kinetic energy along the lattice direction $z$, $a_{0}$ the primary lattice constant, and ${\cal V}_p(z)=V_p\cos^2(k_pz)$ the spin-independent primary lattice realized by a blue-detuned standing wave ${\bf E}_2$. The $t_{\rm so}$ is generated by the Raman potential $t_{\rm so}=\int dzw_s^*(z)\mathcal{M}(z)w_s(z-a_0/2)$,
where ${\cal M}(z)=M_0\cos(k_pz)$ represents the Raman potential formed by the intersection of $x$-polarized ${\bf E}_2$ and $z$-polarized traveling wave ${\bf E}_3$, generating the spin-flipped hopping. The QP on-site matrix element is realized by the spin-dependent secondary lattice, which is given by $M_zV_{j}=\int dz{\cal V}_{s}(z)\abs{w_s(z-ja_0)}^{2}=V_{s}\cos(2\pi\alpha j + 2\phi_s)$,
where the QP parameter $\alpha = k_s / k_p$, and ${\cal V}_s(z) = V_s \cos^2(k_s z + \phi_s)$ is the spin-dependent secondary lattice generated by the composite standing wave ${\bf E}_1$, formed by two counter-propagating beams with mutually perpendicular polarization along the $x$ and $y$ axes. The chiral parameter $\eta$ is tuned by adjusting the power ratio of the two different polarizations of beam ${\bf E}_1$. The spin-dependent term $\eta {\cal V}_s (z)\sigma_z$ from vector optical shift is induced by the y-polarization, while the spin-independent term $(1-\eta) {\cal V}_s (z)\sigma_0$ is mainly contributed from the scalar components of both the $x$- and $y$-polarizations.

{The different phases can be distinguished by the dynamical exponent $\alpha$ governing wave-packet spreading, $\sigma(t)\sim t^{\alpha}$, where $\sigma(t)$ denotes wave packet width. The exponent takes $\alpha=1$ for extended phase, $0<\alpha<1$ for critical phase, and $\alpha=0$ for localized phase. By preparing initial states at different energies and measuring the corresponding expansion dynamics, one can observe energy-dependent (energy-independent) dynamical exponents, providing direct experimental evidence for mobility edges (pure phases).}

\section{Conclusion and outlook}
\label{sec:Conclusion}
We have proposed 
{a broad class of} spin-$1/2$ quasiperiodic (QP) system that
unifies the existing important 1D QP models, some of which requiring
an extra Majorana representation, and established a unified framework for all fundamental localization phases in quasiperiodic systems. The framework is built on three universal results. First, we showed with renormalization
group method the criteria for obtaining the
pure phases: {the mobility edges are absent when the system has chiral(-like) symmetry.} Second, we uncovered a new fundamental mechanism for the
emergence of critical states, as characterized by the generalized incommensurate zeros in matrix elements that remain invariant under dual transformation and introduce critical orbitals into the system. Third, we identified the condition
for the exact solvability of QP models from local constraint, which
also provides the mechanism to realize effective unbounded on-site
potential in spinful QP chains. These results establish a profound unified theory for all fundamental localization phases in quasiperiodic systems, and offer the key insights for constructing their exactly solvable models. In particular, we proposed the novel spin-selective QP lattice model and QP optical Raman lattice model, in which we predict all basic types of mobility edges (MEs) and all the seven fundamental phases of localization physics in QP systems, respectively, which have not been achieved anywhere else. 
The experimentally feasible schemes for realization have been proposed and studied in detail.

This work enables a broad exploration and exact characterization of the fundamental quantum phases in Anderson localization and opens up intriguing topics for the future study. For example, while the present study is focused on the spin-$1/2$ QP systems, it is natural to extend the current study to the QP systems with larger spins, e.g. to the 1D $\mathrm{SU}(N)$ QP chain
with $N$ larger than $2$. Generalizing the mechanism of IDZs in matrix elements for critical states to the $\mathrm{SU}(N)$ systems is of great interests to obtain the rigorous critical states in generic QP systems. The exact characterization of the distinct quantum phases in such $\mathrm{SU}(N)$ QP system could potentially be connected to the exactly solvable QP mosaic models \citep{wang2020a,zhou2023c}
with large unit cells, and deserves particular future efforts.
On the other hand, extending the present rigorous results to {the system with long-range hopping, ladder geometries, and further to} higher dimensions is another important issue for the future study.

Moreover, the newly proposed exactly solvable models in this work, which host
various MEs, can further be implemented in the many-body regime. The different types of the MEs in this work, including the traditional ME separating localized and extended states, and the novel MEs involving critical states, can be considered to study the interplay between many-body localization, ergodic and many-body critical phases. The exact solvability of the model, which allows for the analytic characterization of the whole non-interacting spectrum, serves as an analytic platform to study the interplays among them.

\acknowledgments

This work was supported by National Key Research and Development Program of China (No.~2021YFA1400900, No.~2022YFA1405800 and No.~2020YFA0713300), the National Natural Science Foundation of China (No.~12425401, No.~12261160368, No.~12401208, No.~12531006 and No.~12526201), the Innovation Program for Quantum Science and Technology (No.~2021ZD0302000), the Shanghai Municipal Science and Technology Major Project (No.~2019SHZDZX01), the Natural Science Foundation of Jiangsu Province (No. BK20241431), the Nankai Zhide Foundation, and by the Fundamental Research Funds for the Central Universities (Grant No. 100-63253099).

\appendix

\section{1D spinless QP model} \label{app:Models}

In this section, we show that the 1D spinless QP model with nearest-neighbor hopping can also be unified within the spinful QP framework, under the Majorana representation. For a generic 1D spinless QP chain model nearest-neighbor hopping, the tight binding Hamiltonian is
given by
\begin{equation}
H=t_{j}\sum_{j}(c_{j}^{\dagger}c_{j+1}+\mathrm{h.c.})+\sum_{j}V_{j}c_{j}^{\dagger}c_{j}. \label{eq:spinlessQP}
\end{equation}
The $t_{j}$ and $V_{j}$ are the hopping coefficients and on-site
potential, respectively. Such 1D spinless QP chain can be rewritten into the spinful QP
chain by introducing the Majorana representation
\begin{equation}
	c_{j}=\frac{1}{2}(\gamma_{B,j}+i\gamma_{A,j}), \ c_{j}^{\dagger}=\frac{1}{2}(\gamma_{B,j}-i\gamma_{A,j}),
\end{equation}
with $\gamma_{\sigma,j}=\gamma_{\sigma,j}^{\dagger}$ and $\{\gamma_{\sigma,j},\gamma_{\sigma;,j'}\}=2\delta_{\sigma\sigma'}\delta_{jj'}$. In this basis, the Hamiltonian in Eq.~\ref{eq:spinlessQP} becomes
\begin{eqnarray}
	H &=&\frac{t_{j}}{2}\sum_{j} (i\gamma_{B,j}\gamma_{A,j+1}+ i\gamma_{B,j+1}\gamma_{A,j}+) \nonumber\\
&+&\sum_{j}\frac{V_{j}}{2}(1+i\gamma_{B,j}\gamma_{A,j}).
\end{eqnarray}
One can neglect the last summation, which leads to a constant, then the generic Hamiltonian is unified within our 1D spinful formalism, with the hopping coupling matrix and the on-site matrix given by
\begin{equation}
\Pi_{j}=\frac{t_{j}}{4}\sigma_{y},\ M_{j}=\frac{V_{j}}{4}\sigma_y.
\end{equation}
One characteristic is  that both hopping coupling and on-site matrix share the same Pauli component $\sigma_y$, which is a hallmark of the spinless QP model rewritten in the spinful QP framework. This formalism unifies all the important 1D spinless QP model in this context, including the AA model~\cite{aubry1980}, the extended AA (EAA) model~\cite{hatsugai1990,han1994,liu2015}, the generalized AA (GAA) model~\cite{ganeshan2015}, or the GPD model in some context. These are summarized in the table.~\ref{Table:models}.


\begin{table}[th]
	\centering
	\begin{tabular}{|c|c|c|}
	\hline
	\textbf{Models}  &  $t_{j}$ &  $V_{j}$   \\ \hline
	\text{AA model}  & $t$     &   $ 2V_0 V_j^{\mathrm{d}}$    \\ \hline
	\text{EAA model}  & $(t+\mu)V_j^{\mathrm{od}}$  & $2V_0 V_j^{\mathrm{d}}$   \\ \hline
	\text{GAA model}  & $t$  & $2V_0 V_j^{\mathrm{d}}/(1-aV_j^{\mathrm{d}})$  \\ \hline
	\end{tabular}
	\caption{The parameters of different 1D spinless QP model}
	\label{Table:models}
\end{table}

The conditions for the different states of the AA, EAA and GAA models are summarized in Table~\ref{Table:1DQPPhaseCondition}.
\begin{table}[h!]
	\centering
	\begin{tabular}{|c|c|c|c|}
	\hline
	&\textbf{Extended}&\textbf{Critical} & \textbf{Localized} \\ \hline
	{\rm AA}  &  $t>V_0$  &  $t=V_0$  &  $t<V_0$ \\ \hline
	{\rm EAA} &  $t>{\rm max} \{\mu, V_0 \}$  & $\mu>{\rm max}\{t, V_0 \}$ & $V_0>{\rm max} \{t, \mu \}$   \\ \hline
	{\rm GAA} & $a E<2(t-V_0)$   &  $a E=2(t-V_0)$ & $a E>2(t-V_0)$   \\ \hline
	\end{tabular}
	\caption{The criteria for different states of AA, EAA and GAA models. Here $E$ is the eigenenergies of the system. Without loss of generality, we assume of the hopping and on-site coefficients are positive.}
	\label{Table:1DQPPhaseCondition}
\end{table}

\section{1D QP mosaic model} \label{app:Mosaic}
In this section, we introduce the one-dimensional QP mosaic model~\cite{wang2020a,zhou2023c} and rewrite it into our 1D spinful QP framework. The generic QP mosaic lattice refers that the modulation of on-site potential and hopping coefficients are in a mosaic manner, namely the the QP modulation or the uniform modulation only appears every $\kappa$ sites, as we will see in the following.

The original type-I and type-II QP mosaic model is distinguished by the presence of alternating mosaic pattern in hopping coefficients. If the QP modulation only exhibits in the on-site potential, 
then the mosaic models belongs to type-I, with Hamiltonian ~\cite{wang2020a}
\begin{equation}
H = \lambda \sum_j (c_j^\dagger c_{j+1} + {\rm h.c.}) + 2 \sum_j V_j c_j^\dagger c_{j},
\end{equation}
where the on-site potential is given by
\begin{equation}
V_j =
\begin{cases}
V_0 \cos(2\pi\alpha j + \theta), & j =0 \quad {\rm mod}~\kappa, \\
0, & j \neq 0 \quad {\rm mod}~\kappa.
\end{cases}
\end{equation}
Here $ c_j $ is the annihilation operator at site $ j $, and $ \lambda $, $ V $, and $ \theta $ denote the nearest-neighbor hopping coefficient, on-site potential amplitude, and phase offset, respectively. $ \alpha $ is an irrational number, and $ \kappa $ is an integer. The QP potential periodically occurs every $ \kappa $ sites.

If the system exhibits QP mosaic modulation in hopping coefficients, then it belongs to type-II QP mosaic model, with Hamiltonian~\cite{zhou2023c}
\begin{equation}
	H = \sum_j (t_j c_j^\dagger c_{j+1} + \text{H.c.}) + \sum_j V_j  c_j^\dagger c_{j} ,
\end{equation}
here both the quasiperiodic hopping coefficient $ t_j $ and on-site potential $ V_j $ are in mosaic manners, with the hopping coefficient given by
\begin{equation}
t_j =
\begin{cases}
2t \cos(2\pi\alpha j + \theta), & j = 0 \bmod \kappa, \\
\lambda, & j = 1 \bmod \kappa, \\
\lambda, & \text{otherwise},
\end{cases}
\end{equation}
and the on-site potential given by
\begin{equation}
V_j =
\begin{cases}	
2t \cos(2\pi\alpha j + \theta), & j = 0 \bmod \kappa, \\
2t \cos[2\pi\alpha (j-1) + \theta], & j = 1 \bmod \kappa, \\
0, & \text{otherwise}.
\end{cases}
\end{equation}
The conditions for the different states of the two QP mosaic models are summarized in Table~\ref{Table:MosaicPhaseCondition}, and the MEs separating different types of states can be obtained accordingly. The critical state is absent in type-I model and the extended state is absent in type-II model. The critical states in the Type II QP mosaic model are generated by the IDZs in the hopping coefficients~\cite{zhou2023c}.
\begin{table}[h!]
	\centering
	\begin{tabular}{|c|c|c|c|}
	\hline
	&\textbf{Extended}&\textbf{Critical} & \textbf{Localized} \\ \hline
	{\rm Type I}&$E<|\lambda^2/V_0|$ & \text{none}  &  $E>|\lambda^2/V_0|$ \\ \hline
	{\rm Type II}& \text{none}  &  $E<|\lambda|$ & $E>|\lambda|$   \\ \hline
	\end{tabular}
	\caption{The criteria for different states of type I and type II mosaic models. Here $E$ is the eigenenergies of the system, $\lambda$ and $V_0$ are the coefficients defined in the main text. The critical state is absent in type I model and the extended state is absent in type II model.}
	\label{Table:MosaicPhaseCondition}
\end{table}

If we focus on the case $\kappa =2$, then both two QP mosaic lattice models exhibit a even-odd sublattice structure, which can be rewritten as the spinful QP chain model. We can relabel the even and odd site as spin-up and spin-down degrees of freedom and further relabel the site index, then the Hamiltonian can be rewritten as
\begin{equation}
	H_{\mathrm{M}} = \sum_{j} \lambda(c^{\dagger }_{j+1}\sigma_{-}c_{j}+\mathrm{h.c.})+\sum_{j}c_{j}^{\dagger}V_j^{\mathrm{M}}c_{j},
\end{equation}
where $c_{j}=(c_{j,\uparrow},c_{j,\downarrow})^{\intercal}$ is the
spinor for the annihilation operators. The parameter $V_j^{\mathrm{M}}$ controls the type-I and type-II QP mosaic lattice, which respectively given by
\begin{eqnarray}
	V_j^{\mathrm{M,I}}&=&2 V_0 V_j^{\mathrm{d}}\Lambda_{+}+\lambda\sigma_x,\\
	V_j^{\mathrm{M,II}}&=&2tV_j^{\mathrm{d}}(\sigma_0+\sigma_x).
\end{eqnarray}


\section{Proof of criteria for the pure phases without MEs}\label{app:ProofI}

In this section, we provide the detail of the proof of the criteria
for the system exhibits pure phase and exhibits no mobility edges
(MEs), as discussed in the main text, using the renormalization group
technique.

Under the commensurate approximation of the irrational parameter $\alpha_n=p_n/q_n$, with the size of rational unit cell fixed to $L=q_n$, the system displays periodic structure and band dispersion. And the band dispersion is periodic with respect to both $\kappa_{x}=Lk_{x}$
and $\kappa_{y}=Lk_{y}$. Here, $k_{x}$ is the twisted momentum attached
to the hopping coupling matrix $\Pi_{j}\rightarrow\Pi_{j}e^{ik_{x}}$
and $k_{y}$ is the phase offset in the QP modulation $V_{j}^{\mathrm{d}}$
and $V_{j}^{\mathrm{od}}$.
The characteristic polynomial is given by the determinant $P^{(n)}(E;\kappa_{x},\kappa_{y})=|{\cal H}^{(n)}-E|$,
which can be expanded over the principle dispersion
\begin{align}
P^{(n)}(E;\kappa_{x},\kappa_{y}) & =t_{R}^{(n)}\cos(\kappa_{x}+\kappa_{x}^{0})+V_{R}^{(n)}\cos(\kappa_{y}+\kappa_{y}^{0})\nonumber \\
 & +\mu_{R}^{(n)}\cos(\kappa_{x}+\tilde{\kappa}_{x}^{0})\cos(\kappa_{y}+\tilde{\kappa}_{y}^{0})\nonumber \\
 & +\epsilon_{R}^{(n)}(E,\kappa_x,\kappa_y)+T_{R}^{(n)}(E).
\end{align}
{where $t_R^{(n)}(E)$, $V_R^{(n)}(E)$ and $\mu_R^{(n)}(E)$ are renormalized coefficients of fundamental dispersion $\cos(\kappa_x+\kappa_x^0)$, $\cos(\kappa_y+\kappa_y^0)$ and $\cos(\kappa_x+\tilde{\kappa}_x^0)\cos(\kappa_y+\tilde{\kappa}_y^0)$, respectively. The $\kappa_x^0$, $\kappa_y^0$, $\tilde{\kappa}_x^0$ and $\tilde{\kappa}_y^0$ are model-dependent phase offsets.
The term $\epsilon_{R}^{(n)}$ collects the renormalized contributions from higher harmonic (higher-frequency) modes, while $T_R^{(n)}(E)$ denotes the dispersionless renormalized coupling independent of $\kappa_x$ and $\kappa_y$.}

In the presence of chiral(-like) symmetry, the Hamiltonian excludes  spin-independent ($\sigma_0$) term. We then diagonalize the on-site matrix $M_{j}$ as $M_{j}=m_{j}^{z}\sigma_{z}$, and the resulted hopping coupling matrix retains  the form $\Pi_{j}=\sum_{s}p_{j}^{s}\sigma_{s}$.
When hopping coupling matrices satisfy
\begin{equation}
p_{j}^{x}/p_{j}^{y}=\mathrm{Const.}\in\mathbb{R}, \label{eq:app_chiral}
\end{equation}
then the system exhibits the chiral symmetry. We simplify notation by setting $m_{j}^{z}\equiv m_{j}$. 
The transverse hopping coupling components can be parameterized as
\begin{equation}
p_{j}^{x}=p_{j}^{\bot}\cos\theta,\ p_{j}^{x}=p_{j}^{\bot}\sin\theta,
\end{equation}
with $p_{j}^{\perp}=\sqrt{|p_{j}^{x}|^{2}+|p_{j}^{y}|^{2}}$ and $\tan\theta=p_{j}^{x}/p_{j}^{y}$. Under a unitary
transformation $U(z,\theta)={\rm diag}(e^{i\theta}, e^{-i\theta})$, the eigenvalue equation can be rewritten as
\begin{equation}
\Pi'_{j-1}\vec{v}_{j-1}+\Pi'{}_{j}^{\dagger}\vec{v}_{j+1}+M_{j}\vec{v}_{j}=E\vec{v}_{j},\label{eq:ProofI}
\end{equation}
with the transformed hopping coupling matrix being
\begin{equation}
\Pi'_{j}=p_{j}^{\perp}\sigma_{x}+p_{j}^{z}\sigma_{z},
\end{equation}
thus, under the unitary operator ${\cal O}=\sigma_{y}$, both coupling
matrices anticommute
\begin{equation}
{\cal O}\Pi'_{j}{\cal O}^{\dagger}=-\Pi'_{j},\ {\cal O}M_{j}{\cal O}^{\dagger}=-M_{j}.
\end{equation}
Therefore both $\vec{v}_{j}$ and ${\cal O}\vec{v}_{j}$ are the
eigenstate of the system, with the corresponding energies come in
pair as $(E,-E)$, confirming the chiral(-like) symmetry.

\begin{widetext}
With this properties, we proceed to calculate the characteristic
polynomial $P(E)$. The eigen-equation $H\lvert v\rangle=E\lvert v\rangle$ under the rotation $\psi_{j}^{\pm}=v_{j\uparrow}\pm iv_{j\downarrow}$ is given by
\begin{equation}
\begin{array}{c}
e^{ik_{x}}(p_{j-1}^{z}+ip_{j-1}^{\perp})\psi_{j-1}^{-}+e^{-ik_{x}}(p_{j}^{z*}+ip_{j}^{\perp*})\psi_{j+1}^{-}+m_{j}\psi_{j}^{-}=E\psi_{j}^{+}\\
e^{ik_{x}}(p_{j-1}^{z}-ip_{j-1}^{\perp})\psi_{j-1}^{+}+e^{-ik_{x}}(p_{j}^{z*}-ip_{j}^{\perp*})\psi_{j+1}^{+}+m_{j}\psi_{j}^{+}=E\psi_{j}^{-}
\end{array}
\end{equation}
We further perform the Fourier transformation,
then $P(E)=\det|H-EI|=\det|(A_{mj})(E)_{L \times L}|$.
By denoting the hopping coefficients $t_{j}^{\pm}=p_{j}^{z}\pm ip_{j}^{\perp}$, each block being a 2-by-2 matrix is given by
	\begin{equation}
		A_{mj}=e^{i2\pi\alpha jm}\left(\begin{array}{cc}
		t_{j-1}^{+}e^{-i(2\pi\alpha m+k_{x})}+t_{j}^{-*}e^{i(2\pi\alpha m+k_{x})}+m_{j} & -E\\
		-E & t_{j-1}^{-}e^{-i(2\pi\alpha m+k_{x})}+t_{j}^{+*}e^{i(2\pi\alpha m+k_{x})}+m_{j}
		\end{array}\right)
		\end{equation}
\end{widetext}
In the following, we show that the dominant dispersions 
have energy-independent coefficients.

We first prove the coefficients that give rise to extended and localized
states are energy independent. We first investigate the parameters
attached to the leading dispersion $\cos Lk_{x}$ and $\cos Lk_{y}$.
Since each term within the determinant is obtained by a multiply of
$2L$ terms, the terms consisting $\cos Lk_{x}$ comes from the multiplication
of $L$ folds $e^{ik_{x}}$ and $L$ folds energies $E$. This indicates
that such terms are illegal under the chiral symmetry since such term
is of odd power of energy $\sim E^{L}\cos Lk_{x}$, which is also
the same for $\cos k_{y}$. So the parameters associated with dispersion
$\cos Lk_{x}$ and $\cos Lk_{y}$ vanish. Then the coefficients that
give rise to extended and localized states, become the parameters
along with the dispersion $\cos2Lk_{x}$ and $\cos2Lk_{y}$. These
two terms can only be obtained by multiply all the diagonal terms
in $\det A$, therefore they are energy independent.

Now we prove the parameters associated with the critical states, which
are linked to the dispersion $\cos n_{x}Lk_{x}\cos n_{y}Lk_{y}$ with
$n_{x},n_{y}\geq1\in\mathbf{\mathbb{Z}}$, are also energy independent.
Since we are considering the case $p_{j}^{x}/p_{j}^{y}=\mathrm{Const.}\in\mathbb{R}$
and $M_{j}$ are purely quasiperiodic, there ate basically two choices
for the form of the hopping coupling matrix $\Pi_{j}$: One is that the
hopping coupling matrix $\Pi_{j}$ are uniform while the on-site matrix $M_{j}$ are purely quasiperiodic. And the other choice is that
both $\Pi_{j}$ and $M_{j}$ are purely quasiperiodic.

We first prove that for the case $\Pi_{j}$ are uniform, the coefficient
associated with $\cos Lk_{x}\cos Lk_{y}$ is energy independent. This
is because the dispersion $\cos Lk_{x}\cos Lk_{y}$ comes from the
product of combination of $L$ folds $p_{j}$ terms and combination
of $L$ folds $m_{j}$ terms, which contribute to $L$ folds $e^{ik_{x}}$
and $e^{ik_{y}}$, respectively. Then the total multiplication of
$2L$ elements contribute the $\cos Lk_{x}\cos Lk_{y}$, leaving no
chance for involving $E$.

To formalize this argument, we analyze the determinant structure of the Fourier-transformed matrix $A(E)$. For uniform $\Pi_j$, the leading momentum-dependent contributions to $\det A$ can be systematically extracted by expanding the determinant over permutations and retaining only the lowest-frequency harmonics.

When $\Pi_j$ is uniform, the leading coefficients can be obtained explicitly by isolating the dominant contributions in the permutation expansion of $\det A$. In general,
\begin{equation}
\det A
= \sum_{\sigma \in S_{2L}} \mathrm{sgn}(\sigma)
\prod_{m,s} A_{m s;\sigma_{m s}},
\end{equation}
where $S_{2L}$ is the permutation group of of indexes $1,\ldots,2L$, and $s=\pm1$ labels internal indices within each block $A_{mj}$. The relevant coefficients arise from contributions of the form
\begin{equation}
\sum_{\sigma}
\mathrm{sgn}(\sigma)
\prod_{m,s}
\exp\!\left(
i 2\pi \alpha m s \sigma_{m s}
\right),
\end{equation}
multiplied by the following momentum-dependent factors
\begin{align}
\Big\{&
\Big[
\Big(\prod_m t^+ e^{-i(2\pi\alpha m+k_x)}\Big)
\Big(\prod_j m^z \cos(2\pi\alpha j+k_y)\Big)
+ \mathrm{h.c.}
\Big]
\nonumber\\
&+ (t^+ \leftrightarrow t^{-*})
\nonumber\\
&+
\Big[
\Big(\prod_m t^+ e^{-i(2\pi\alpha m+k_x)}\Big)
\Big(\prod_m t^- e^{-i(2\pi\alpha m+k_x)}\Big)
+ \mathrm{h.c.}
\Big]
\nonumber\\
&+
\Big[\prod_j m^z \cos(2\pi\alpha j+k_y)\Big]^2
\Big\}.
\end{align}
Retaining only the lowest-frequency harmonics, the determinant reduces to
\begin{align}
&\det A
\simeq \nonumber\\ \xi_0 \Big[
&
\xi_C
\Big(\frac{m^z}{2}\Big)^L
(|t^+|^L + |t^-|^L)
\cos(L k_x + \tilde{\kappa}_x^0)
\cos(L k_y)
\nonumber\\
+&
\xi_T |t^+ t^-|^L
\cos(2L k_x + \kappa_x^0)
\nonumber
+
\xi_V
\Big(\frac{m^z}{2}\Big)^{2L}
\cos(2L k_y)
\Big]\nonumber\\
+& \text{(higher-frequency terms)} .
\end{align}
Here the common prefactor $\xi_0$ is given by
\begin{equation}
\xi_0
=
\sum_{\sigma}
\mathrm{sgn}(\sigma)
\prod_{m,s}
\exp\!\left(
i 2\pi \alpha m s \sigma_{m s}
\right),
\end{equation}
and plays no role in the scaling analysis. The coefficients $\xi_C$, $\xi_T$, and $\xi_V$ are nonuniversal constants of order unity, while the phases $\kappa_x^0$ and $\tilde{\kappa}_x^0$ depend solely on $\arg(t^\pm)$. From this expression, the effective renormalized coefficients for the leading dispersions can be directly obtained
\begin{align}
t_{2R}^{(L)} &\approx |t^+ t^-|^L ,\\
V_{2R}^{(L)} &\approx \Big(\frac{m^z}{2}\Big)^{2L} ,\\
C_R^{(L)} &\approx
\Big(\frac{m^z}{2}\Big)^L
\big(|t^+|^L + |t^-|^L\big) .
\end{align}
These results align with phase boundary in Table.~\ref{Table:1DSOC}~\cite{wang2020b,goncalves2023a}. In this derivation, we assume that intermediate power combinations, such as $(t^+)^{L-1}(t^-)^{L+1}$, do not contribute at leading order. Such terms enter the permutation sum with effectively random phases and  cancel out, leading to a parametric suppression relative to the coherent lowest-harmonic terms.

Then we consider the case both $\Pi_{j}$ and $M_{j}$ are purely QP,
in which case the leading dispersion is 
$\cos Lk_{x}\cos2Lk_{y}$ or $\cos2Lk_{x}\cos2Lk_{y}$, whose coefficients
are all energy independent. A key ingredient is that the hopping coupling
matrix exhibits the off-diagonal modulation $(V_{j}^{\mathrm{od}}c_{j+1s}^{\dagger}c_{js'}+\mathrm{h.c.})$,
which makes the $p_{j}$ contributes the $k_{x}$ and $k_{y}$ simultaneously.
Therefore, if $\Pi_{j}$ are QP, $\cos Lk_{x}\cos Lk_{y}$
dispersion will disappear. The reason is that $\cos Lk_{x}\cos Lk_{y}$
comes from the multiplication of $L$ folds $p_{j}$ terms and $L$
folds $E$ terms, which is not allowed under the chiral(-like) symmetry.
So the leading order contributing to the critical orbital in this
case is $\cos Lk_{x}\cos2Lk_{y}$ or $\cos2Lk_{x}\cos Lk_{y}$. The
first case corresponds multiplication of $L$ folds $p_{j}$ terms
and $L$ folds $m_{j}$ terms, which are in total $2L$ terms and
leaves no chance to involve the extra terms. And the latter dispersion cannot be realized in the system. So the coefficient associated
with $\cos Lk_{x}\cos 2Lk_{y}$ is energy independent. Similarly,
the $\cos2Lk_{x}\cos 2Lk_{y}$ originates from the product of $2L$
folds $p_{j}$ terms and is also energy independent.

\section{Generalized 2D bilayer mapping}\label{app:DualMapping}
In this section, we provide the detail of the mapping the spinful QP model to a generalized 2D bilayer square lattice system, in the presence of an external magnetic field with an irrational flux threading each unit cell. We provide the detail that the uniform and QP hopping couplings in $\Pi_j$ are mapped to the hopping couplings along $x$-direction and diagonal directions, respectively. And the uniform and QP on-site coupling in $M_j$ are mapped to on-site coupling and hopping along $y$-direction, respectively. This mapping is done by interpreting the phase shift $k_{y}$ as the quasi-momentum in the y-direction and the spin index  $s$ as the layer index. Performing a Fourier transformation, the coefficient $V_{j}^{\rm od}$ in hopping coupling matrix $\Pi_{j}$ and $V_{j}^{\rm d}$ in the on-site matrix $M_{j}$ contribute the hopping terms along diagonal and y-directions,respectively.  In the following, we analyze this generalized mapping of different processes.

Under the 2D mapping, the uniform hopping in Hamiltonian is mapped to the intra-layer (for $s=s'$) or inter-layer (for $s\neq s'$) hopping coupling along $x$-direction
\begin{equation}
t^{ss'} (c_{j+1 s}^{\dagger} c_{j s^{\prime}}+{\rm h.c.})\rightarrow t^{ss'}(a_{x+1,y,s}^{\dagger}a_{x,y,s'}+{\rm h.c.}),
\end{equation}
with $t^{s s'}$ being the uniform hopping coefficients of different elements in $\Pi_{j}$. The QP hopping term in the Hamiltonian is mapped to the intra-layer or inter-layer hopping coupling in the diagonal direction in the presence of magnetic flux, which is given by
\begin{eqnarray}
\mu^{s s'} V_{j}^{\rm od}( c_{j+1 s}^{\dagger} c_{j s^{\prime}}+{\rm h.c.})&\rightarrow& \frac{\mu^{s s'}}{2}(e^{-i 2\pi\alpha x} a_{x+1, y+1, s}^{\dagger} a_{x, y, s^{\prime}} \nonumber\\
	& +&e^{i 2\pi\alpha x} a_{x+1, y-1, s}^{\dagger} a_{x, y, s'}\nonumber\\
& +&{\rm h.c.}),
\end{eqnarray}
where $\mu^{s s'}$ are the QP hopping coefficients. The uniform on-site coupling $\lambda^{s s^{\prime}} c_{j s}^{\dagger} c_{j s^{\prime}}$ is mapped to the uniform onsite term or inter-layer coupling. For $s=s^{\prime}$, the mapping yields the on-site energy shift, and when $s\neq s^{\prime}$, it gives the inter-layer coupling
\begin{equation}
\lambda^{s s^{\prime}} c_{j s}^{\dagger} c_{j s^{\prime}}\rightarrow
\lambda^{ss^{\prime}}(a_{x,y,s}^{\dagger}a_{x,y,s^{\prime}}+{\rm h.c.}).
\end{equation}
Finally the QP on-site coupling $V^{s s^{\prime}} V_{j}^{\rm d} c_{j s}^{\dagger} c_{j s^{\prime}}$ characterizes the intra-layer or inter-layer hopping coupling in the $y$-direction under the magnetic field, with $V^{s s^{\prime}}$ being the hopping coefficients
\begin{eqnarray}
V^{s s^{\prime}} V_{j}^{\rm d} c_{j s}^{\dagger} c_{j s^{\prime}}\rightarrow\frac{V^{ss^{\prime}}}{2}(e^{-i2\pi\alpha x}a_{x,y+1,s}^{\dagger}a_{x,y,s^{\prime}}+{\rm h.c.}).\nonumber\\
\end{eqnarray}
From the mapping given above, we can see that the dominant constant hopping term (quasiperiodic onsite term) in the original 1D model gives the dominant hopping along the $x$ ($y$) direction in the mapped 2D bilayer system, and the extended (localized) states emerge. In contrast, when the quasiperiodic hopping dominates in the 1D model, in the 2D bilayer system the inter-layer diagonal hopping coupling dominates over other terms. In this case the states are extended in the diagonal direction, hence being delocalized in both $x$ and $y$ directions. 

The states delocalized in both directions are critical states, and are invariant under dual transformation. In the mapped 2D bilayer system, the dual transformation $c_{j,s}=(1/\sqrt{L})\sum_{n}e^{-i2\pi n\alpha j}c_{n,s}$ is equivalent to a gauge transformation, which is given by
\begin{equation}
	a^{\dagger}_{x,y,s}\rightarrow e^{i\varphi xy}a^{\dagger}_{x,y,s}  ,\ \vec{A}=(-\varphi x,0) \rightarrow \vec{A}=(0,\varphi y).
\end{equation}
The resulted Hamiltonian after the transformation yields
\begin{align}
	H =&\sum_{x,y,s,s'}\Big[ t^{ss^{\prime}}e^{i \varphi y}a_{x+1,y,s}^{\dagger}a_{x,y,s^{\prime}} +\frac{V^{ss^{\prime}}}{2}a_{x,y+1,s}^{\dagger}a_{x,y,s^{\prime}} \nonumber \\
	&+\lambda^{ss^{\prime}}a_{x,y,s}^{\dagger}a_{x,y,s^{\prime}}+ \frac{\mu^{s s^{\prime}}}{2}(e^{i \varphi y} a_{x+1, y+1, s}^{\dagger} a_{x, y, s^{\prime}}  \nonumber \\
	& +e^{-i \varphi y} a_{x+1, y-1, s}^{\dagger} a_{x, y, s^{\prime}})   + {\rm h.c.} \Big]. \label{eq:2DspinHam_dual}
\end{align}
It is seen that the diagonal hopping terms and on-site coupling terms in the 2D bilayer remain invariant under the dual transformation, while the hopping along the $x$ and $y$ directions are interchanged. Thus when the hopping along diagonal direction dominates, while those along $x$ and $y$ directions are irrelevant, the states are self-dual and are critical.

\section{Exact solvability of QP systems from local constraint} \label{app:ExactSolvable}


In this section, we provide the details of the exact solvability of the spinful QP lattice model under the local constraint $\det|\Pi_{j}|=0$ or $\det|\Pi_{n}|=0$.

We begin by briefly reviewing Avila's global theory for one-frequency QP Schr\"odinger operators~\cite{avila2015}. Let $T$ be an analytic function mapping the circle $S^{1}$ into the group $SL(2,\mathbb{C})$.
An analytic cocycle $(\alpha,T)$ is defined as a linear skew product
\begin{equation*}
(\alpha,T): S^{1}\times\mathbb{R}^{2}\rightarrow S^{1}\times\mathbb{R}^{2},\qquad
(\theta,\psi)\mapsto (\theta+\alpha, T(\theta)\,\psi).
\end{equation*}
If $T(\theta)$ admits a holomorphic extension to  $|\Im\theta|<\delta$, then for $|\epsilon|<\delta$ we define the complexified cocycle
$T_\epsilon(\theta)=T(\theta+i\epsilon)$
and the associated LE
\begin{equation}
\gamma_\epsilon(T)=\lim_{n\to\infty}\frac{1}{n}\int \ln\|T_n(\theta+i\epsilon)\|\,d\theta,
\end{equation}
where $T_n$ denotes the transfer matrix at site $n$.

Imposing the exact-solvability condition, the original spinful QP lattice model can be rigorously reduced to an effective 1D tight-binding problem with NN hopping $t^{\rm eff}_j$ and onsite potential $V^{\rm eff}_j$. The resulting transfer matrix inherits a constrained analytic structure, allowing the LE to be evaluated explicitly via Avila's global theory.



\subsection{New exactly solvable mosaic models}
We present the analytic derivation for the LE of the QP spin-flipped (QPSF) model in Eq.~\eqref{eq:QPSF}. The effective hopping and onsite potential are given by
\begin{equation*}
t_j^{\rm eff}=2\lambda t \cos(2\pi\alpha j),\quad
V_j^{\rm eff}=\lambda^2+4t^2\cos^2(2\pi\alpha j).
\end{equation*}	
The corresponding complexified transfer matrix can be written in the factorized form $T_j(\theta + i\epsilon) = \tilde{T}_j(\theta + i\epsilon) / (2\cos(\theta + \alpha/2 + i\epsilon))$, where the numerator matrix $\tilde{T}_j$ has elements $\tilde{T}_j^{11}=[E^2 -\lambda^2-4t^2\cos^2(\theta+i\epsilon)]/\lambda t$, $\tilde{T}_j^{12}=-2 \cos(\theta-\alpha+i\epsilon)$, $\tilde{T}_j^{21}=2\cos(\theta+i\epsilon)$, and $\tilde{T}_j^{22}=0$.  The explicit factor $2\cos(\theta)$, which corresponds to the IDZs discussed in the main text, immediately excludes absolutely continuous spectral components and therefore forbids extended states. Taking the asymptotic limit $\epsilon\to\infty$, one finds $\tilde{T}_j=e^{-i\theta+2\epsilon}[T' + \mathcal{O}(e^{-\epsilon})]$, where the only nonvanishing element of $T'$ is $T'_{11}=-t/\lambda$.

Now we calculate the LE. Applying Jensen's formula, we obtain
\begin{equation}
	\int_{0}^{2 \pi}\ln |2\cos(\theta+i\epsilon)| {\mathrm{d} \theta}=2\pi |\epsilon|.
\end{equation}
The total complexified LE is therefore
\begin{eqnarray*}
	\begin{split}
2\gamma_\epsilon 
 =&\lim _{m \to \infty} \frac{1}{2 \pi m} \int_{0}^{2 \pi} \ln \left\|\tilde{T}_{j+m-1} \cdots \tilde{T}_{j}\right\| \mathrm{d} \theta \\
&-\int_{0}^{2 \pi}\ln |2\cos(\theta+\alpha/2+i\epsilon)| {\mathrm{d} \theta}\\
 = & \ln \| T' \| +2|\epsilon|-|\epsilon| +\mathcal{O}\left(\epsilon^{-1}\right) \\
 =& \ln |t/\lambda| + |\epsilon| +\mathcal{O}\left(\epsilon^{-1}\right) \\
	\end{split}.
\end{eqnarray*}
Since $2\gamma_\epsilon(E)$ is a convex, piecewise-linear function of $\epsilon$ with quantized slopes, this asymptotic behavior uniquely fixes the Lyapunov exponent for all $\epsilon\in\mathbb{R}$ as $2\gamma_\epsilon = \max \left\{0, \ln |t/\lambda| + |\epsilon|\right\}$. Setting $\epsilon=0$, we obtain
\begin{equation}
\gamma = \max \left\{0, \frac{1}{2}\ln |t/\lambda|\right\},
\end{equation}
Which is the LE in the main text. A positive LE corresponds to the localized phase, and for the zero LE, it corresponds to the critical states. Finally, applying the duality transformation yields the phase diagram of the corresponding dual mosaic model in the main text.

\subsection{The spin-selective QP lattice model}
We procede to give the details of the SSQP model. As discussed in the main text, the SSQP model and its dualcounterpart are exactly solvable at the condition $t=V_0=V_B=0$ and $\lambda_0=\lambda_1=\lambda$, for which the effective hopping and onsite potential reduce to the forms given in Eq.~\eqref{eq:TeffGIZs}. Following the same strategy as above, we decompose the complexified transfer matrix as $T_j(\theta + i\epsilon) = \tilde{T}_j(\theta + i\epsilon) / (2\cos(\theta + \alpha/2 + i\epsilon))$, with
\begin{equation*}
\tilde{T}_j=
\begin{pmatrix}
\frac{E}{\mu} - \frac{\lambda^2(4 + 4\cos(\theta + i\epsilon))}{\mu E} & -2\cos(\theta - \alpha/2 + i\epsilon) \\
2\cos(\theta + \alpha/2 + i\epsilon) & 0
\end{pmatrix},
\end{equation*}
Taking the limit $\epsilon\to\infty$, we obtain $\tilde{T}_j=e^{-i\theta+\epsilon}[T' + \mathcal{O}(e^{-\epsilon})]$, with the $T'$ given by
\begin{equation}
T'
=
\begin{pmatrix}
{2\lambda^2}/{\mu E} & -e^{i\alpha/2} \\
e^{-i\alpha/2} & 0
\end{pmatrix}
\end{equation}
Combining with Jenson's formula, the LE takes the form
\begin{eqnarray*}
	\begin{split}
2\gamma_\epsilon(E) 
 =&\lim _{m \to \infty} \frac{1}{2 \pi m} \int_{0}^{2 \pi} \ln \left\|\tilde{T}_{j+m-1} \cdots \tilde{T}_{j}\right\| \mathrm{d} \theta \\
&-\frac{1}{2\pi}\int_{0}^{2 \pi}\ln |2\cos(\theta+\alpha/2+i\epsilon)| {\mathrm{d} \theta}\\
 = & \ln \| T' \| +\mathcal{O}\left(\epsilon^{-1}\right) \\
	\end{split}
\end{eqnarray*}
Since $2\gamma\epsilon(E)$ is a convex, piecewise-linear function of $\epsilon$ with quantized slopes, this asymptotic behavior uniquely determines
\begin{equation}
\gamma_\epsilon(E)=\frac{1}{2}  \max \left\{0,\ln\Big| \big|\lambda^{2}/\mu E\big|+\sqrt{(\lambda^{2}/\mu E)^{2}-1}\Big|\right\}.
\end{equation}
Setting $\epsilon=0$ recovers the Lyapunov exponent quoted in Eq.~\eqref{eq:LE_SSQPi} of the main text.

We next provide analytic details for the SSQP model in the case $\mu=\lambda_1=0$, where the effective hopping and onsite potential are given in Eq.~\eqref{eq:TVeffGIZsII}.
The transfer matrix can be factorized as $T_j=T_j^{A}T_j^{B}$, with $T_j^A=t/(E-V_0-2V_BV_j^d)$
and
\begin{equation*}
T_j^{B}=
\begin{pmatrix}
\frac{E(E-V_0-2V_BV_j^d)}{t^2}-\lambda_0^2 & -\frac{E-V_0-2V_BV_j^d}{t} \\
\frac{E-V_0-2V_BV_j^d}{t} & 0
\end{pmatrix},
\end{equation*}
Accordingly, the LE decomposes as $2\gamma=\gamma_A+\gamma_B$.
To evaluate $\gamma_B$, we complexify $\theta\to\theta+i\epsilon$ and take $\epsilon\to\infty$, yielding $T_j^B=e^{-i\theta+\epsilon}[\mathcal{T}^B+\mathcal{O}(e^{-\epsilon})]$, with  $\mathcal{T}^B_{11} = -{E V_B}/{t^2}$, $\mathcal{T}^B_{12} = {V_B}/{t}$, $\mathcal{T}^B_{21} = -{V_B}/{t}$, and $\mathcal{T}^B_{22} = 0$. And the complexified $\gamma_B$ up to order $\mathcal{O}(\epsilon^{-1})$ is given by
\begin{equation*}
\gamma_B=\ln \Big|\frac{|E V_B/t^2|+\sqrt{(E V_B/t^2)^2-(2V_B/t)^2}}{2}\Big|+2\pi|\epsilon|.
\end{equation*}
Now we calculate $\gamma_A$ under different parameter regimes. When $|E-V_0|>2V_B$, one finds
\begin{equation*}
	\begin{split}
\gamma_A=&\int_0^{2\pi}\ln\Big| \frac{t}{E-V_0-2V_B\cos(\theta+i\epsilon)}\Big|d\theta \\
=&\ln\Big|\frac{t}{V_B}\Big|-2\pi|\epsilon|,
	\end{split}
\end{equation*}
Therefore we have the total Lyapunov exponent given by
\begin{equation}
\gamma(E)=\ln\Big|\frac{|E/t|+\sqrt{(E/t)^{2}-4}}{2}\Big|,
\end{equation}
which gives the LE in the main text Eq.~\eqref{LE1}. When $|E-V_0|<2V_B$, the  $\gamma_A$ is given by
\begin{equation*}
	\begin{split}
\gamma_A=&\int_0^{2\pi}\ln\Big| \frac{t}{E-V_0-2V_B\cos(\theta+i\epsilon)}\Big|d\theta \\
=&\ln \left| \frac{2}{| \chi_0 | + \sqrt{\chi_0^2 - \chi_B^2}} \right|,
	\end{split}	
\end{equation*}
and the total Lyapunov exponent becomes
\begin{equation}
	\gamma(E)=\ln\Bigg|\frac{|\chi_{E}|+\sqrt{\chi_{E}^{2}-\chi_{B}^{2}}}{|\chi_{0}|+\sqrt{\chi_{0}^{2}-\chi_{B}^{2}}}\Bigg|+2\pi|\epsilon|,
\end{equation}
where  $\chi_{B}=2V_{B}/t$, $\chi_{E}=EV_{B}/t^{2}$ and $\chi_{0}=(E-V_{0})/t$. Setting $\epsilon=0$ reproduces Eq.~\eqref{LE2} of the main text.

\renewcommand{\thesection}{S-\arabic{section}}
\setcounter{section}{0}  
\renewcommand{\theequation}{S\arabic{equation}}
\setcounter{equation}{0}  
\renewcommand{\thefigure}{S\arabic{figure}}
\setcounter{figure}{0}  
\renewcommand{\thetable}{S\Roman{table}}
\setcounter{table}{0}  
\onecolumngrid \flushbottom 

\newpage
\begin{center}\large \textbf{Supplementary Material} \end{center}

\section{Semiclassical approximation of the critical states} \label{app:Semiclassical}

We take the semiclassical approximation to study how the critical states are obtained under the mechanisms of the generalized incommensurate zeros (GIZs) in hopping coupling matrix $\Pi_{j}$ and the incommensurately distributed zeros (IDZs) in the matrix component of on-site matrix $M_{j}$ shared in $\Pi_{j}$. Note that while this is an approximate study, it provides an intuitive picture for the emergence of the critical states. 

\begin{figure}[ht]
	\includegraphics{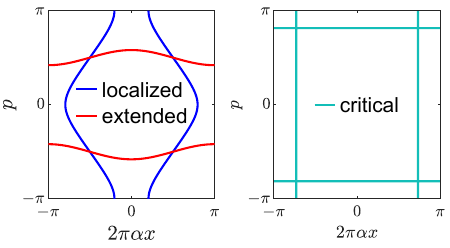}
	\caption{\label{fig:semiclass} Phase space trajectories for extended, localized and critical states. The extended state is continuous in $x$ direction while the localized state is continuous along momentum direction. The critical state is continuous in both directions. Here the extended, localized and critical states are realized in the extended AA model by setting $(t,V,\mu)=(1,2.5,0)$, $(t,V,\mu)=(1,0.5,0)$ and $(t,V,\mu)=(1,2.5,1.5)$, respectively.}
\end{figure}
The semiclassical description is well valid for small irrational number $\alpha \ll 1$, in which case the QP modulation varies slowly across the lattice. We first take the example of the extended AA model introduced in Appendix~\ref{app:Models}, which supports all three types of states. Then we have the energy in the phase space
\begin{equation}
E (p,x) = [2 t + 2 \mu \cos (2 \pi \alpha x)] \cos p + V \cos (2 \pi
\alpha x).
\end{equation}
Here $x=j$ labels the site coordinate and $p$ is the momentum, so that the contours $E(p,x)=E$ define the phase-space trajectories~\cite{Albert2010,Malla2018} of the extended AAH model. As depicted in Fig.~\ref{fig:semiclass}, the extended  and localized states correspond to the trajectories continuous in $x$ direction but discontinuous in $p$ direction, whereas localized states exhibit the opposite behavior. The critical states are continuous in both directions. The dual transformation here is swapping $2\pi x$ and $p$. One can notice that $2 \mu \cos (2 \pi \alpha x)\cos p$ remains invariant under the dual transformation and always percolates in both directions, responsible for the critical states.

The semiclassical phase-space trajectory for the current spin-$1/2$ QP system decomposes into two branches,
\begin{equation}
E_{\pm}(p,x)
  = h_{0} \pm \sqrt{h_{x}^{2} + h_{y}^{2} + h_{z}^{2}}, \label{eq:semiclassSpin}
\end{equation}
where each component is given by
\begin{align}
	h_{0} & =V_{0}\cos\bigl(2\pi\alpha x\bigr)
		+ 2t_{0}\cos p
		+ 2\mu_{0}\cos\bigl(2\pi\alpha x\bigr)\cos p,\\
	h_{x}& = \lambda_{x}
		+ V_{x}\cos\bigl(2\pi\alpha x\bigr)
		+ 2 |t_{x}|\cos\bigl(p + \phi_{t_x}\bigr) \nonumber \\
			&+ 2|\mu_{x}|\cos\bigl(2\pi\alpha x\bigr)\cos\bigl(p + \phi_{\mu_x}\bigr),\\
	h_{y} &=\lambda_{y}
		+ V_{y}\cos\bigl(2\pi\alpha x\bigr)
		+ 2 |t_{y}|\cos\bigl(p + \phi_{t_y}\bigr) \nonumber \\
			&+ 2|\mu_{y}|\cos\bigl(2\pi\alpha x\bigr)\cos\bigl(p + \phi_{\mu_y}\bigr),\\
	h_{z}& =\lambda_{z}
		+ V_{z}\cos\bigl(2\pi\alpha x\bigr)
		+ 2t_{z}\cos p \nonumber \\
		&+ 2\mu_{z}\cos\bigl(2\pi\alpha x\bigr)\cos p.
\end{align}	
Here, $|t_{x,y}|$ and $\phi_{t_{x,y}}, \phi_{\mu_{x,y}}$ denote the magnitudes and phase angles of the generally complex  spin-flipped hopping coefficients, while the spin-conserved hopping coefficients are taken to be real up to a global gauge choice, and all the other parameters are real.

The critical states generated from the GIZs can be understood as the phase-space trajectory percolation due to explicit dual invariant terms. When the system exhibits GIZs in the hopping coupling matrix $\Pi_{j}$, arising from either the spin-flipped or spin-conserved QP terms, the $\cos(2\pi\alpha x)\cos p$ contribution dominates the local dispersion, yielding a percolating, dual-invariant trajectory in phase space and thus producing critical eigenstates.

The semiclassical picture also explains that the IDZs in the matrix component of $M_j$ shared in $\Pi_j$ generate critical states. Without losing generality, we consider IDZs in $\sigma_z$ component of $M_j$ shared in $\Pi_j$  as an example. Specifically, the uniform hopping couples to $\sigma_z$ component and $M_j$ exhibits IDZs in $\sigma_z$ component. This case has no GIZs for $\mu_{0,x,y,z}=0$, the $h_z$ takes the generic form
\begin{equation}
	h_z = \lambda_z + V_z\cos\bigl(2\pi\alpha x\bigr) + 2t_z\cos p,
\end{equation}
where $t_z$ is the uniform hopping coefficient, and $V_z$ is the QP on-site coupling, contributing IDZs in $\sigma_z$ component. The $h_z^2$ in Eq.~\eqref{eq:semiclassSpin} produces the cross term $V_z\,2t_z\cos\bigl(2\pi\alpha x\bigr)\cos p$, which gives rise to the critical orbitals in the system. One can see that the IDZs in the matrix component of $M_j$ shared in $\Pi_j$ leads to the percolated phase-space trajectory, which is a mechanism unique to spinful QP systems.

\section{Evolution of phase boundaries away from exactly solvable limits} \label{app:PhaseBoundary}

  Upon departing from the exactly solvable limits, the phase boundaries generically evolve smoothly in parameter space; however, whether the spectrum develops a complicated structure of mobility edges (MEs) depends crucially on the symmetries preserved in the corresponding parameter regime. In particular, chiral symmetry can forbid MEs altogether, leading to smooth deformations of phase boundaries, whereas broken chiral symmetry allows energy-dependent restructuring of the spectrum and the emergence of intricate ME structures.

In this section, we show that upon departing from the exactly solvable limits, the phase boundaries generically evolve smoothly in parameter space; however, whether the spectrum develops a complicated structure of mobility edges (MEs) depends crucially on the symmetries preserved in the corresponding parameter regime. In particular, chiral symmetry can forbid MEs altogether, leading to smooth deformations of phase boundaries, whereas broken chiral symmetry allows energy-dependent restructuring of the spectrum and the emergence of intricate ME structures.

\begin{figure*}
        \centering
        \includegraphics[scale=0.95]{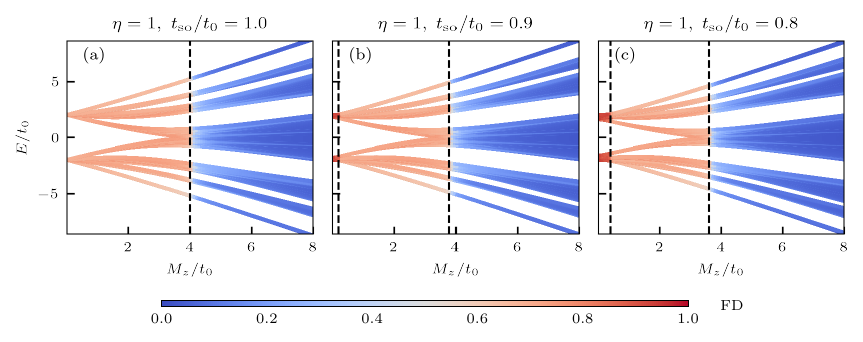}
        \caption{Departing from exactly solvable limits. Phase diagrams at fixed $\eta=1$ illustrating smooth deformations away from the exactly solvable point. The fractal dimension (FD) as functions of $M_z/t_0$ and energy $E/t_0$ for $t_{\rm so}/t_0=1.0$ (a), $0.9$ (b), and $0.8$ (c), respectively. The dashed vertical lines mark the analytically predicted phase boundaries $M_z=2|t_0\pm t_{\rm so}|$. Despite the imbalance between $t_0$ and $t_{\rm so}$, the absence of mobility edges persists due to the chiral(-like) symmetry, and the phase boundaries shift continuously.}
        \label{fig:FD_eta1_varying_tso}

    \end{figure*}

    When the chiral symmetry is preserved, MEs are absent and the phase boundaries deform smoothly upon departing from the exactly solvable point. This behavior is illustrated in Fig.~\ref{fig:FD_eta1_varying_tso}, where we introduce an imbalance between $t_0$ and $t_{\rm so}$ while keeping $\eta=1$.
    In this case, the phase boundaries between critical-localized and critical-extended phases shift continuously according to $M_z=2|t_0\pm t_{\rm so}|$.


    \begin{figure*}
        \centering
        \includegraphics[scale=0.95]{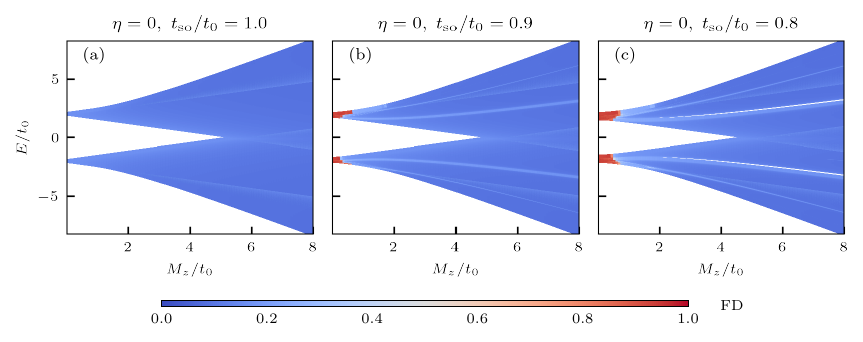}
        \caption{Departing from exactly solvable limits. Phase diagrams at fixed $\eta=0$ illustrating the emergence of mobility edges away from the exactly solvable point. The fractal dimension (FD) as functions of $M_z/t_0$ and energy $E/t_0$ for $t_{\rm so}/t_0=1.0$ (a), $0.9$ (b), and $0.8$ (c), respectively. Upon introducing an imbalance between $t_0$ and $t_{\rm so}$, extended states appear at small $M_z$ and energy-dependent mobility edges develop.}
        \label{fig:FD_eta0_varying_tso}
    \end{figure*}

    \begin{figure*}
        \centering
        \includegraphics[scale=0.95]{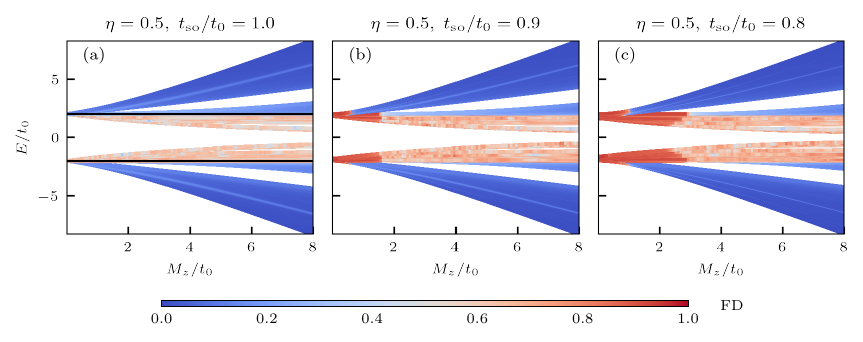}
        \caption{Departing from exactly solvable limits. Phase diagrams at fixed $\eta=0.5$ showing the development of mobility edges. The fractal dimension (FD) as functions of $M_z/t_0$ and energy $E/t_0$ for $t_{\rm so}/t_0=1.0$ (a), $0.9$ (b), and $0.8$ (c), respectively. In contrast to the chiral-symmetric case, the imbalance between $t_0$ and $t_{\rm so}$ induces energy-dependent effective hopping processes (see main text), leading to extended states and nontrivial mobility edges.}

        \label{fig:FD_eta05_varying_tso}

    \end{figure*}

\begin{figure*}
        \centering
        \includegraphics[scale=0.95]{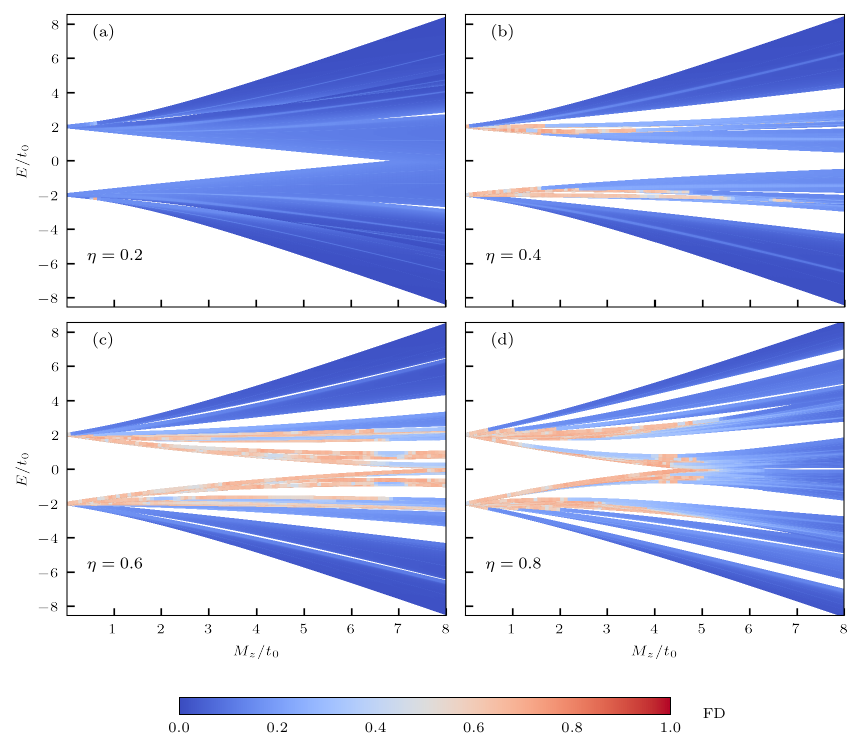}
        \caption{Departing from exactly solvable limits. Phase diagrams illustrating the evolution of mobility edge upon varying $\eta$ at fixed $t_{\rm so}=t_0$. The fractal dimension (FD) as functions of $M_z/t_0$ and energy $E/t_0$ for $\eta=0.2$ (a), $0.4$ (b), $0.6$ (c), and $0.8$ (d), respectively. Away from the exactly solvable points $\eta=0,\,0.5,\,1$, $\eta\neq1$ breaks chiral symmetry and introduces energy dependent quasiperiodic hopping processes, leading to critical states and associated mobility edges.}

        \label{fig:FD_tso1_varying_eta}

    \end{figure*}

In contrast, when the chiral symmetry is broken, the phase diagram can develop a more intricate structure of MEs. Specifically, for $\eta=0$ and $\eta=0.5$, as shown in Figs.~\ref{fig:FD_eta0_varying_tso} and \ref{fig:FD_eta05_varying_tso}, relaxing the condition $t_0=t_{\rm so}$ leads to the MEs associated with extended states. This behavior can be attributed to the effective next-nearest-neighbor hopping induced by the imbalance between $t_0$ and $t_{\rm so}$, which has an energy-dependent form
\begin{equation}
	t_j^{\rm NN}=\frac{t_{+}t_{-}}{E-\Delta_{j+1}},
\end{equation}
where $t_{\pm}=t_0\pm t_{\rm so}$. This process favors extended states at small $M_z$ but depends nontrivially on energy, thereby producing complex ME structures. At sufficiently large $M_z$, these additional processes become subdominant and the overall phase diagram remains qualitatively unchanged.

A similarly rich ME structure also arises when departing from exact solvability by varying $\eta$ while keeping $t_{\rm so}=t_0$. In this case, the model lacks chiral symmetry, and a finite $\eta$ simultaneously introduces energy dependent IDZs and higher-order (quartic) contributions in $M_z$, leading to energy-dependent localization transitions. As shown in Fig.~\ref{fig:FD_tso1_varying_eta}, varying $\eta$ from 0 to 1 interpolates between the three exactly solvable points $\eta=0, 0.5, 1$, during which complicated MEs generically emerge. This behavior can be understood from the effective quasiperiodic hopping
\begin{equation}
	t_{j}^{\rm eff}=\frac{-2t_0\eta\Delta_j}{E-(1-\eta \Delta_j )},
\end{equation}
which induces critical states, together with the effective onsite potential
\begin{equation}
	V_{j}^{\mathrm{eff}}=\frac{4t_{0}^{2}}{E-(1-\eta)\Delta_{j-1}}+\frac{E(1-\eta)\Delta_{j}-(1-2\eta)\Delta_{j}^{2}}{E-(1-\eta)\Delta_{j}}.
\end{equation}
For $0<\eta<0.5$ and $0.5<\eta<1$, the coexistence of single- and double-frequency quasiperiodic modulations renders the locations of MEs highly nontrivial.

\section{Experimental Scheme} \label{app:ExpScheme}
In this section, we provide a more detailed description of the experimental scheme for realizing the Hamiltonian Eq.~\eqref{eq:GIZs_Exp} in the main text. Without loss of generality, we consider $\ket{\uparrow}\equiv\ket{F=1,m_F=-1}$ and $\ket{\downarrow}\equiv\ket{1,0}$ of $^{87}{\rm Rb}$ atoms as an example, while all results remain applicable to other alkali atoms. After applying the unitary transformation $\sigma_z\rightarrow\sigma_x$ and $\sigma_x\rightarrow-\sigma_z$, the target Hamiltonian  can be rewritten as
\begin{align}
	H=\left[\frac{p_z^2}{2m}+{\cal V}_s(z)\right]\otimes\sigma_0+{\cal V}_p(z)(\sigma_0-\sigma_x)/2-M_0\sigma_z,
\end{align}
where ${\cal V}_p(z)$ and ${\cal V}_s(z)$ represent the primary and secondary optical lattices, respectively. These lattices are generated by two sets of standing waves and a periodic Raman potential. To construct the secondary lattice, we require a standing wave of the form
\begin{align}
	{\bf E}_1 = 2E_1 \hat{e}_x e^{{\rm i}(\phi_1 + \phi_1^\prime / 2)} \cos(k_1 z - \phi_1^\prime / 2),
\end{align} where $\phi_1$ is the initial phase of the incident beam, and $\phi_1^\prime$ is an additional phase acquired by ${\bf E}_1$ before being reflected back to the atoms. Given that the modulus of the wave vector is
$k_1=2\pi/\lambda_1$, and ignoring the overall constant energy shift, the standing wave field ${\bf E}_1$ induces a spin-independent potential given by
\begin{equation}
	{\cal V}_1(z) = \frac{V_1}{2} \cos(2k_1 z - \phi_1^\prime),\quad V_1 = \sum_{J,F^\prime,m_F^\prime} \frac{\abs{\Omega_{F^\prime,m_F^\prime;\sigma,10}^{(J)}}^2}{\Delta_{F^\prime,m_F^\prime}^{(J)}}.
\end{equation}

\begin{figure}[!t]
	\includegraphics[width=0.5\columnwidth]{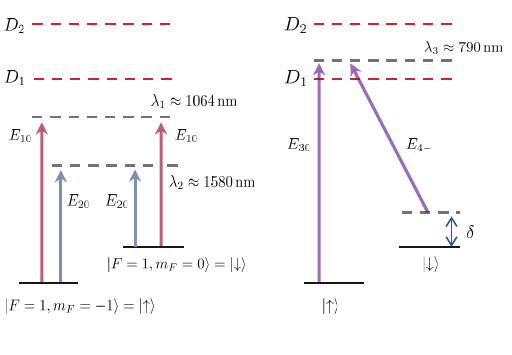}
	\caption{\label{fig:SM_Figs} The proposed laser configuration enables the realization of a model featuring all three fundamental types of mobility edges. Laser beams ${\bf E}_1$ and ${\bf E}_2$ with wavelengths $\lambda_1$ and $\lambda_2$ ,individually generate spin-independent lattices, while ${\bf E}_3$ and ${\bf E}_4$ operating at `tune-out' wavelength $\lambda_3$ form a periodic Raman potential $(\,^{87}\text{Rb atoms})$.}
\end{figure}

In the above, the transition matrix elements are $\Omega_{F^\prime,m_F^\prime;\sigma,nq}^{(J)}=-\langle F^\prime,m_F^\prime|{\bf d}|\sigma\rangle E_n/\hbar$ and $\Delta_{F^\prime,m_F^\prime}^{(J)}$ are detunings of laser field from different levels. Here, $J=1/2$, $3/2$ correspond to the $D_1$ and $D_2$ lines respectively, and $\sigma=\uparrow, \downarrow$ labels the spin states. The index $n$ denotes the laser beam, while $q=0, \pm$ represents the polarization of laser field. For a laser wavelength of $\lambda_{1} \approx 1064\,{\rm nm}$, the overall optical shifts induced by the linearly polarized ($q=0$) field from ${\bf E}_1$ are nearly identical for the two spin states. Comparing this with Eq.~\eqref{eq:GIZs_Exp}, we find $k_s=2k_1,\ V_s=V_1/2$ and $\phi_s=-\phi_1^\prime$.
On the other hand, the primary lattice ${\cal V}_p(z)$ is introduced through three laser fields, given by
\begin{align}
{\bf E}_2 &= 2E_2 \hat{e}_x e^{{\rm i}(\phi_2 + \frac{\phi_2^\prime}{2})} \cos(k_2 z - \frac{\phi_2^\prime}{2}), \nonumber \\
{\bf E}_3 &= 2E_3 \hat{e}_x e^{{\rm i}(\phi_3 + \frac{\phi_3^\prime}{2})} \cos(k_3 z - \frac{\phi_3^\prime}{2}), \nonumber \\
{\bf E}_4 &= E_4 \hat{e}_z e^{-{\rm i}k_3 y + {\rm i}\phi_4}.
\end{align}
We first define the spin-independent part of the primary lattice
${\cal V}_p(z)$ as a periodic potential generated by the standing wave ${\bf E}_2$, given by
\begin{equation}
	{\cal V}_p(z) = \frac{V_2}{2} \cos(2k_2z - \phi_2^\prime), \label{eq:SM_exp} \quad V_2 = \sum_{J,F^\prime,m_F^\prime} \frac{\abs{\Omega_{F^\prime,m_F^\prime; \sigma, 20}^{(J)}}^2}{\Delta_{F^\prime, m_F^\prime}^{(J)}}.
\end{equation}
Then the standing wave ${\bf E}_3$ and the traveling wave ${\bf E}_4$ together generate a periodic Raman potential along the z-direction
\begin{equation}
	{\cal V}_p(z) = M_{34} e^{{\rm i}(\phi_4 - \phi_3 - \phi_3^\prime/2)} \cos(k_3 z - \phi_3^\prime / 2),
\end{equation}
with $M_{34}$ given by
\begin{equation}
	M_{34} = \sum_{J, F^\prime, m_F^\prime} \frac{\Omega_{F^\prime, m_F^\prime; \uparrow, 30}^{(J)*} \Omega_{F^\prime, m_F^\prime; \downarrow, 4-}^{(J)}}{\sqrt{2} \Delta_{F^\prime, m_F^\prime}^{(J)}}.
\end{equation}
Here, we adopt the following gauge convention for the spherical basis $\hat{e}_z=-1(\hat{e}_+-\hat{e}_-)/\sqrt{2}$ and $\hat{e}_y={\rm i}(\hat{e}_++\hat{e}_-)/\sqrt{2}$. For simplicity, we choose `tune-out' wavelength $\lambda_3 \approx 790\,{\rm nm}$~\cite{Wen2021}. As a result, the lattice depth generated by ${\bf E}_3$ is nearly zero, since the optical shifts from the $D_1$ and $D_2$ lines cancel out~\cite{footnote_Exp}. By comparing this expression with Eq.~\eqref{eq:SM_exp}, we have $k_3=2k_2,\ M_{34}=V_2/2$ and $\phi_3^\prime/2=\phi_2^\prime$.
The exponential phase factor $\phi_4-\phi_3-\phi_2^\prime=0$ can be stabilized by locking the relative phase $\phi_4-\phi_{3}$ using a phase-locking technique~\cite{Jotzu2014,Gorg2019,wang2021}. The two-photon detuning of Raman coupling is defined as $\delta/2=(\Delta E+\omega_3-\omega_4)/2\equiv M_0$, where $\Delta E$ is the energy splitting between $\ket{\uparrow}$ and $\ket{\downarrow}$, typically on the order of tens of MHz and $\omega_{3,4}$ denote the corresponding frequencies of ${\bf E}_{3,4}$. The sources for ${\bf E}_2$ and ${\bf E}_{3,4}$ can be generated using a Raman fiber laser with second-harmonic generation (SHG) cavity, which provides suitable wavelengths of both $\lambda_2$ and $\lambda_{3,4}$ at the same time.  The additional phase $\phi_3^\prime$ and $\phi_2^\prime$ acquired along the reflective path can be finely tuned by varying the optical path length of the reflected beam. Finally, we see that the QD lattice in Eq.~\eqref{eq:GIZs_Exp} can be realized using the proposed scheme when $k_p = 2k_2, \ V_p = V_2/2$ and $\phi_p = -\phi_2^\prime$ are satisfied.

This experimental scheme leads to a ratio of $\alpha=k_p/k_s\approx 0.67$. The nearest-neighbor hopping amplitude $t$, determined by the lattice potential $V_0$, approximately follows the relation $t\approx4V_0^{3/4}e^{-2\sqrt{V_0}}/\sqrt{\pi}$~\cite{li2017}. To suppress the nearest-neighbor hopping for spin-down atoms, the parameters  $V_s=2.5,\,V_p=10,\,M_0=1.5$ are chosen such that the hopping amplitude $t$ for spin-up atoms is about one order of magnitude larger than that for spin-down atoms.


\begin{thebibliography}{130}
\bibitem {footnote_dispersion}
	The dispersion $\cos Lk_x \cos Lk_y$ vanishes due to QP hopping in the hopping coupling matrix, which ensures that $p_j$ contributes $e^{ik_x}e^{ik_y}$ . This dispersion arises from the product of $L$ factors of $p_{j}$
	and $E$, which gives $E^L\cos Lk_x \cos Lk_y$, prohibited by chiral(-like) symmetry. And the dispersion $\cos 2Lk_x \cos Lk_y$ cannot be realized in the system due to the current constriant of the coupling parameters.
\bibitem{footnote_chiral}
		Breaking the chiral(-like) symmetry does
		not necessarily result in mobility edges (MEs). While it enables the
		possibility of MEs, their existence depends on whether MEs fall within the
		eigenstates, which is determined by the microscopic details. In other words,
		even if chiral(-like) symmetry is broken, the system remains in a pure phase as long
		as the MEs do not intersect the spectrum.
\end{thebibliography}
\end{document}